\documentclass[epj-spec]{svjour}
\usepackage{graphicx}
\usepackage{amsfonts}
\usepackage{amssymb}
\begin{document}
\title{Aspects of stochastic resonance in reaction--diffusion
systems: The nonequilibrium-potential approach}
\author{Horacio S. Wio\inst{1}\fnmsep
\thanks{\email{wio@ifca.unican.es}}\and Roberto R. Deza\inst{2}}
\institute{Instituto de F\'{\i}sica de Cantabria, Universidad de
Cantabria and CSIC\\
E-39005 Santander, Spain \and Departamento de F\'{\i}sica, FCEyN,
Universidad Nacional de Mar del Plata\\
De\'an Funes 3350, 7600 Mar del Plata, Argentina}
\abstract{We analyze several aspects of the phenomenon of
stochastic resonance in reaction--diffusion systems, exploiting the
nonequilibrium potential's framework. The generalization of this
formalism (sketched in the appendix) to \emph{extended} systems is
first carried out in the context of a simplified scalar model, for
which stationary patterns can be found analytically. We first show
how \emph{system-size stochastic resonance} arises naturally in
this framework, and then how the phenomenon of \emph{array-enhanced
stochastic resonance} can be further enhanced by letting the
diffusion coefficient depend on the field. A yet less trivial
generalization is exemplified by a stylized version of the
FitzHugh--Nagumo system, a paradigm of the
\emph{activator--inhibitor} class. After discussing for this system
the second aspect enumerated above, we derive from it---through an
adiabatic-like elimination of the inhibitor field---an effective
scalar model that includes a \emph{nonlocal} contribution. Studying
the role played by the \emph{range} of the nonlocal kernel and its
effect on stochastic resonance, we find an optimal range that
maximizes the system's response.}
\maketitle
\section{Introduction}\label{sec:1}
\emph{Stochastic resonance} (SR) is nowadays a paradigm of the
constructive effects of fluctuations on nonlinear systems
\cite{RMP,RPP}. Sketchily, the phenomenon occurs whenever the
Kramers' rate for the transition between attractors matches the
typical frequency of a signal which is incapable by itself to
trigger that transition (i.e. it is \emph{subthreshold}). Whereas
several measures of SR can be defined [the \emph{signal-to-noise
ratio} (SNR) and the \emph{spectral amplification factor} (SAF)
being the main ones], theoretical analysis is usually carried on in
terms of the \emph{two-state approximation} \cite{RMP}. Since its
discovery a quarter of century ago---and besides exploring related
phenomena like e.g., \emph{coherence resonance}
\cite{CR_excitable}---interest has gradually shifted towards
increasingly complex systems, \emph{networks} and \emph{nonlinear
media} being the main directions. Instances of this trend are the
experiments carried out to explore the role of SR in sensory and
other biological functions \cite{biol}, and experiments in chemical
systems \cite{sch}.

Our concern throughout this review will be with nonlinear media
that can be described as \emph{reaction--diffusion} (RD) systems,
namely those that can be thought of as a collection of diffusively
coupled nonlinear units. The possibility of enhancing the system's
response through the coupling of those units
\cite{extend1,otros,extend2,extend2b,quasi,extend3a,extend3c} has
been among the issues explored during the last decade, together
with the ``naturalness'' problem (how does nature manage to make
the system's response less dependent on a fine tuning of the noise
intensity) or that of searching for different ways to control the
phenomenon \cite{claudio,nos3}.

In dissipative dynamical systems, the very notion of
\emph{Lyapunov's function} is as useful as that of attractor
itself. Even when often it cannot be explicitly computed (because
integrability conditions are not readily met), it allows for
picturesque reasoning in terms of ``energy landscapes'' or
``attraction basins''. When those dynamical systems are submitted
to forces that can be modeled (\`a la Langevin) as stochastic, a
new meaning---statistical in nature, akin to the notion of ``free
energy''---is added to the picture (in fact, it is worth mentioning
that the first function known to have the Lyapunov property was
Boltzmann's \(H\)-function). Moreover, even for vanishing noise
intensity, the very existence of stochastic terms (a ``transport
matrix'') can render the system well conditioned regarding
integrability conditions. That was the rationale behind the
definition of \emph{nonequilibrium potential} (NEP)
\cite{GR,IO,IO1,IO2}, two approaches to which are described in the
Appendix. Such NEP is a special Lyapunov's function of the
associated deterministic system, which for nonequilibrium
systems plays a role similar to that played by a thermodynamic
potential in equilibrium thermodynamics \cite{GR}. It is closely
related to the stationary solution of the system's Fokker--Planck
equation, and characterizes the global properties of the dynamics:
attractors, linear and \emph{relative stability} of these
attractors, \emph{height of the barriers} separating attraction
basins. In addition, it allows to evaluate the \emph{transition
rates} among the different attractors \cite{GR,IO,IO1,IO2,WeNew}.
Regarding the problem of SR in extended systems, it was shown that
the knowledge of the NEP allows to obtain a rather complete picture
of the behavior of the output \emph{signal-to-noise ratio} (SNR).
The novelty with nonequilibrium extended systems is that even
pointlike attractors in the medium's infinite-dimensional
phase-space can be nontrivial field configurations (real-space
\emph{patterns}).

In a series of recent papers we have studied the SR phenomenon for
the transitions between two different patterns
\cite{extend2,extend2b,quasi,extend3a,extend3c,extend3b},
exploiting the concept of nonequilibrium potential. In this review
we discuss some recent results concerning different aspects of SR
in RD systems. In Sec.\ \ref{sec:2} we discuss the
phenomenon of \emph{system-size stochastic resonance} (SSSR), and
show how can it be analyzed and understood within a NEP framework
\cite{SSSR7}. In Sec.\ \ref{sec:3}, after reviewing a recent
study on the enhancement of the SNR found for a scalar system with
density-dependent diffusivity \cite{extend3c}, we discuss its
extension \cite{TW} to an array of FitzHugh--Nagumo units
\cite{FHN}. In Sec.\ \ref{sec:4}---through an adiabatic-like
elimination of the inhibitor field in an activator--inhibitor
system---an effective scalar system with a \emph{nonlocal} term is
derived, and the role of the local and nonlocal interactions on the
SR response studied. The main conclusions are finally summarized in Sec.\ \ref{sec:5}.
\section{System-size stochastic resonance}\label{sec:2}
Recent studies on biological models of the Hodgkin--Huxley type
\cite{SSSR1,SSSR2} have shown that ion concentrations along cell
membranes display intrinsic SR-like phenomena as the number of ion
channels is varied. A related result \cite{SSSR3} shows that even
in the absence of external forcing, the regularity of the
collective firing of a set of coupled excitable FitzHugh--Nagumo
units is optimal for a given number of elements. From a physics
point of view, the same phenomenon---called \emph{system-size
stochastic resonance} (SSSR)---has also been found in an Ising
model as well as in a set of globally coupled units described by a
\(\phi^4\) theory \cite{SSSR4}. It has been even shown to arise in
opinion formation models \cite{SSSR5}.

Since the SSSR phenomenon is peculiar to extended systems, there is
an obvious interest in describing it within a NEP framework, that
offers a very general framework for the study of the dependence of
SR and related phenomena on any of the system's parameters. Here we
discuss in some detail a one-component (``scalar'') RD system
\cite{SSSR7}, and briefly refer to other cases analyzed in
\cite{SSSR4} and \cite{SSSR6}.
\subsection{Review of the scalar model}\label{ssec:2.1}
For one-variable dynamical systems, the Lyapunov's function can
always be found by quadrature. This property can be readily
translated to scalar RD systems: the \emph{Lyapunov's functional}
\(\mathcal{F}[\phi]\) fulfills the ``potential'' condition
\(\partial_t\phi(y,t)=-\delta\mathcal{F}[\phi]/\delta\phi(y,t)\),
where \(\delta/\delta\phi(y,t)\) indicates a functional derivative.
This is also the NEP for a \emph{scalar} transport matrix (i.e.\ a
multiple of the unit matrix in the medium's infinite-dimensional
phase space).

The specific model we shall focus on here has a piecewise linear
reaction term, that mimics general bistable RD models \cite{FHN},
e.g.\ those with a cubic-like reaction term. In the following, we
shall exploit some of the results on the influence of general
(partially reflective or \emph{albedo}) boundary conditions found
in \cite{SW}, as well as previous studies of the NEP \cite{IO} and
of SR \cite{extend2,extend2b,extend3a,extend3b}. The particular
dimensionless form of the deterministic model we start with is
\cite{SW,extend2,extend2b}
\begin{equation}\label{eq:1}
\frac{\partial\phi}{\partial t}=D\,\frac{\partial^2\phi}
{\partial y^2}-\phi+\phi_h\,\Theta(\phi-\phi_c),
\end{equation}
where \(\Theta(z)\) is the Heaviside step function, \(\phi_c\) is
the value at which the piecewise-linear ``reaction term'' has the
jump, and \(D\) is a \emph{phenomenological} ``diffusion
coefficient'', not necessarily related to \(\gamma\) in Sec.\
\ref{ssec:2.2} [in Graham's approach, Eq.\ (\ref{eq:15}),
\(D/(\Delta y)^2\) would be the matrix elements \(Q^{\nu\nu\pm1}\)
in a discretization of the Laplacian]. All the effects of the
parameters that keep the system away from equilibrium (such as the
electric current in electrothermal devices like the ballast
resistor \cite{FHN,SW}, or some external reactant concentration in
chemical models) are now included in \(\phi_c\). For the system to
display a bistable behavior, it must be \(0<\phi_c<\phi_h\).

We consider here the class of static structures \(\phi(y)\) studied
in \cite{SW}. They are even solutions to the \emph{stationary}
(\(\partial_t\phi=0\)) version of Eq.\ (\ref{eq:1}) in the
bounded domain \([-y_L,y_L]\), with \emph{equal} albedo boundary
conditions (b.c.) at both ends \[\left.\frac{\partial\phi(y,t)}
{\partial y}\right|_{y=\pm y_L}=\mp k\,\phi(\pm y_L,t).\]
\(k>0\) is called the albedo parameter: the limit \(k\rightarrow0\)
yieds Neumann's b.c.\ \(\partial_y\phi(y,t)|_{y=\pm y_L}=0\) and
the \(k\rightarrow\infty\) one, Dirichlet's b.c.\
\(\phi(\pm y_L,t)=0\).

The explicit form of these static patterns is
\begin{equation}\label{eq:2}
\phi(y)=\phi_h\left\{\begin{array}{lll}
\quad\sinh(y_c)\rho'\left(k,\frac{y_L+y}{\sqrt{D}}\right)/
\rho\left(k,\frac{y_L}{\sqrt{D}}\right)&\qquad&-y_L\leq y\leq-y_c,
\\
1-\cosh(y)\rho\left(k,\frac{y_L-y_c^\pm}{\sqrt{D}}\right)/
\rho\left(k,\frac{y_L}{\sqrt{D}}\right)&\qquad&-y_c\leq y\leq y_c,
\\
\quad\sinh(y_c)\rho'\left(k,\frac{y_L-y}{\sqrt{D}}\right)/
\rho\left(k,\frac{y_L}{\sqrt{D}}\right)&\qquad&\quad y_c\leq y\leq
y_L,\\
\end{array}\right.
\end{equation}
where \(\rho(k,\zeta)=\sinh(\zeta)+k\,\cosh(\zeta)\) and
\(\rho'(k,\zeta)=\partial\rho/\partial\zeta\). The coordinate
values \(y_c^\pm\) at which \(\phi(y_c)=\phi_c\) are
\begin{equation}\label{eq:3}
y_c^\pm=\frac{1}{2}\left[y_L-\ln\left(\frac{Z\pm\sqrt{Z^2+1-k^2}}
{1+k}\right)\right]\quad
\mbox{with}\quad Z=\left(1-\frac{2\phi_c}{\phi_h}\right)\rho
\left(k,\frac{y_L}{\sqrt{D}}\right).
\end{equation}
Each real solution \(y_c^\pm<y_L\) to Eq.\ (\ref{eq:3})
represents a structure with a central ``activated'' zone
(\(\phi>\phi_c\)) and two lateral ``resting'' regions
(\(\phi<\phi_c\)). Figure 5 in \cite{SW} displays the relation
\(y_c/y_L\) \emph{vs} \(k\), for several values of
\(\phi_c/\phi_h\).

\begin{figure}
\centering
\resizebox{.5\columnwidth}{!}
{\includegraphics[bb= 0pt 0pt 204pt 143pt]{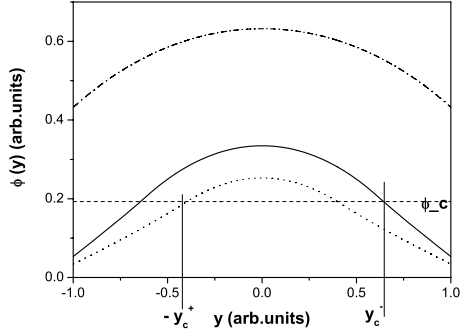}}
\caption{Inhomogeneous static solutions to Eq.\ (\ref{eq:1}) for
\(y_L=D=\phi_h=1.0\) and \(\phi_c=0.193\) (highlighted by the
dashed horizontal line). Dash-dotted upper curve: \(\phi_1(y)\) for
\(k=1.0\) (in this case, \(y_c\equiv y_c^->y_L\)). Lower curves:
\(\phi_1(y)\) (solid line, for which \(+y_c^-\) is highlighted)
and \(\phi_u(y)\) (dotted line, for which \(-y_c^+\) is
highlighted) for \(k=7.0\).}\label{fig:1}
\end{figure}

Typical shapes of the arising patterns are shown in Fig.\
\ref{fig:1}. Through a linear stability analysis it has been shown
\cite{SW} that the structure with the smallest ``excited'' region
[that is with \(y_c=y_c^+\), denoted by \(\phi_u(y)\)] is unstable,
whereas the other one [with \(y_c=y_c^-\), denoted by
\(\phi_1(y)\)] is linearly stable. The trivial homogeneous solution
\(\phi_0(y)=0\) exists and is linearly stable for any parameter
set. These two linearly stable solutions (\(\phi_0\) and
\(\phi_1\)) are the only stable static structures under albedo b.c.
We will concentrate on the region of values of \(\phi_c/\phi_h\),
\(y_L\) and \(k\) where \(\phi_1\) exists.

For the finite system with albedo b.c., the NEP is a functional of
\(\phi\) and a function of \(k\), \(y_L\) and \(\phi_c/\phi_h\). It has the expression
\cite{IO}
\begin{eqnarray*}
&&\mathcal{F}([\phi],\phi_c/\phi_h,k,y_L)=\\
&&\int_{-y_L}^{y_L}\left\{-\int_0^{\phi(y,t)}\left[-\phi'+\phi_h\,
\Theta(\phi'-\phi_c)\right]d\phi'+\frac{D}{2}
\left[\frac{\partial\phi(y,t)}{\partial y}\right]^2\right\}dy+
\left.\frac{k}{2}\,\phi^2(y,t)\right|_{\pm y_L}.
\end{eqnarray*}

When the NEP is evaluated at the \emph{inhomogeneous} static
solutions of Eq.\ (\ref{eq:1}) [Eqs.\ (\ref{eq:2}) and
(\ref{eq:3})] it takes the explicit form \cite{extend2,IO}
\begin{eqnarray}
\mathcal{F}^{u,1}(\phi_c/\phi_h,k,y_L)
&=&\mathcal{F}([\phi_{u,1}],\phi_c/\phi_h,k,y_L)\label{eq:4}\\
&=&\phi_h^2\left[-y_c^\pm\left(1-\frac{2\phi_c}{\phi_h}\right)+
\sinh\left(y_c^\pm/\sqrt{D}\right)
\frac{\rho\left(k,(y_L-y_c^\pm)/\sqrt{D}\right)}
{\rho\left(k,y_L/\sqrt{D}\right)}\right],\nonumber
\end{eqnarray}
while at the trivial solution \(\phi_0\equiv0\) it is
\(\mathcal{F}([\phi_0],y_L)=\mathcal{F}^0=0\).

\begin{figure}
\centering
\resizebox{.45\columnwidth}{!}{\rotatebox{-90}
{\includegraphics[bb= 100pt 0pt 500pt 842pt]{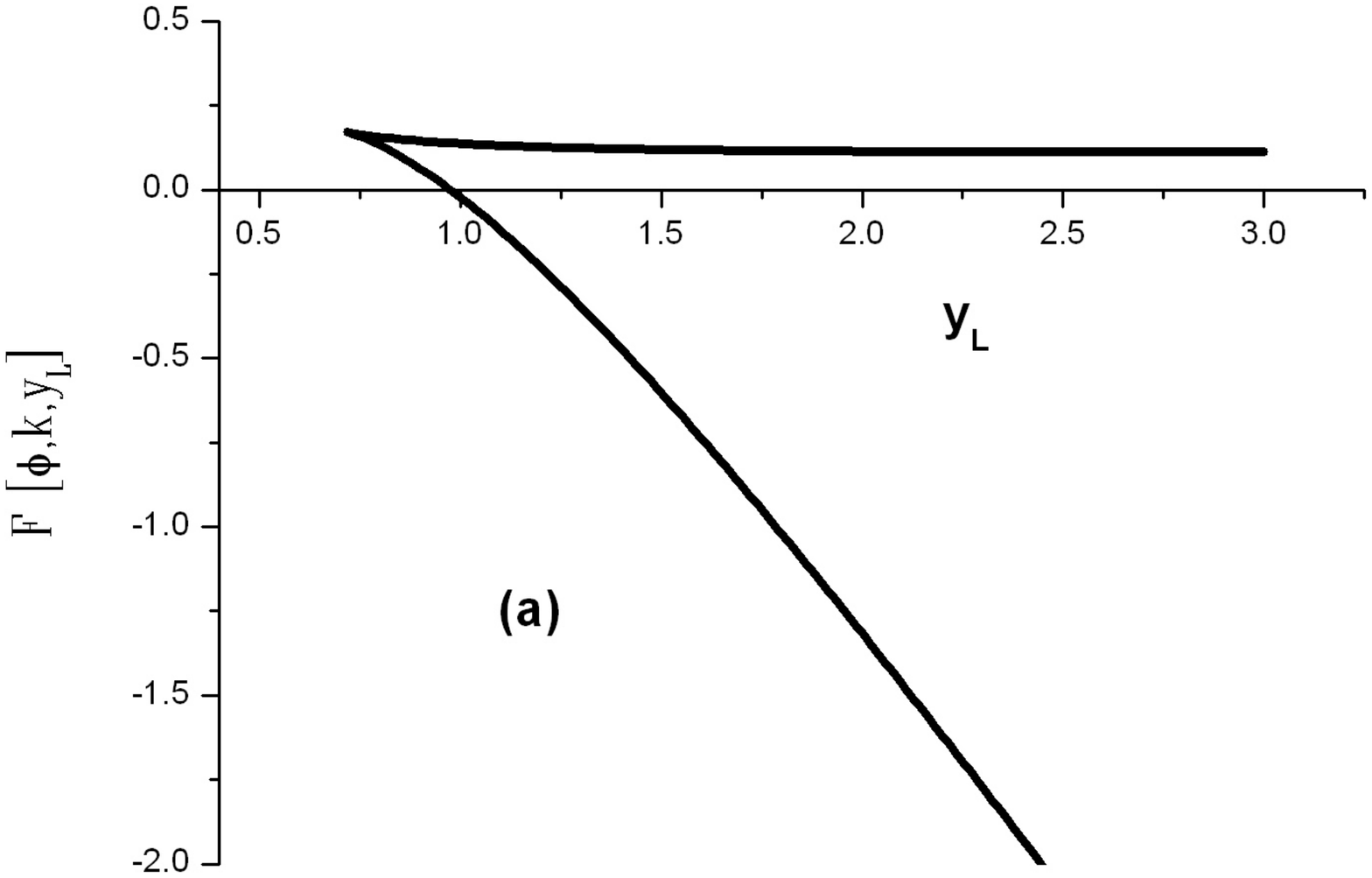}}}
\resizebox{.45\columnwidth}{!}{\rotatebox{-90}
{\includegraphics[bb= 100pt 0pt 500pt 842pt]{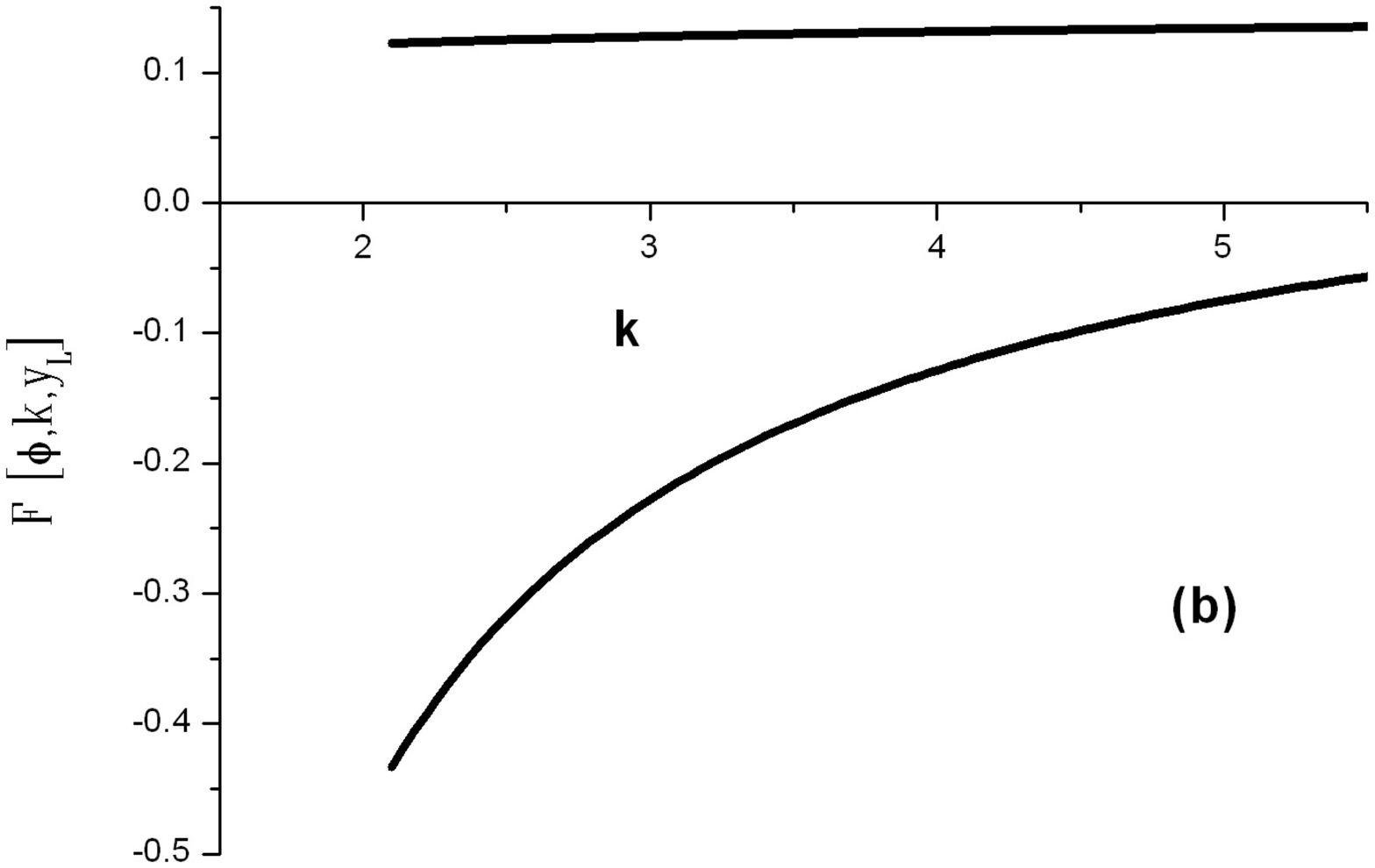}}}
\caption{NEP \(\mathcal{F}([\phi],k,y_L)\) evaluated at the
inhomogeneous stationary solutions \(\phi_1(y)\) (lower branch) and
\(\phi_u(y)\) (upper branch), as a function of: \textbf{(a)}
system's size \(y_L\), with \(k=3.0\); \textbf{(b)}  albedo
parameter \(k\), with \(y_L=1.2\). The remaining parameters are
\(D=1.0\), \(\phi_c/\phi_h=0.193\). The NEP for the homogeneous
stationary solution \(\phi_0(y)\) coincides with the horizontal
axis.}\label{fig:2}
\end{figure}

Figure \ref{fig:2}a depicts the nonequilibrium potential
\(\mathcal{F}([\phi],y_L)\) as a function of the system's size
\(y_L\), keeping the albedo parameter \(k\) and the ratio
\(\phi_c/\phi_h\) fixed. The curves correspond to the NEP
\(\mathcal{F}^{u,1}(y_L)\), whereas \(\mathcal{F}^0\) coincides
with the \(x\)--axis. Our focus is the ``bistable zone''
\(y_L\gtrapprox0.72\), where \(\phi_1(y)\) exists. The unstable
structure \(\phi_u(y)\) is a \emph{saddle point} for
\(\mathcal{F}[\phi]\) (in the medium's infinite-dimensional
phase space), so its NEP \(\mathcal{F}^u(y_L)>0\) (upper branch in
Fig.\ \ref{fig:2}a). On the other hand, both \(\phi_1(y)\) (lower
branch) and \(\phi_0\) (\(x\)--axis) are local minima of the NEP.

One immediately notices that \(\Delta\mathcal{F}^1(y_L)
\equiv\mathcal{F}^u(y_L)-\mathcal{F}^1(y_L)\) is an (almost
linearly) \emph{increasing} function of \(y_L\) (this has a
profound implication for SSSR, as we shall see below). Equation
(\ref{eq:3}) has real solutions only for \(y_L\gtrapprox0.72\).
This corresponds to a \emph{supercritical} saddle--node
bifurcation, at which both inhomogeneous structures pop up. Now,
the most important feature in Fig.\ \ref{fig:2}a is that
\(\mathcal{F}^1(y_L)\) \emph{vanishes} at a certain system's
size \(y_L^*\) (\(\approx1.0\) for the given values of \(k\) and
\(\phi_c/\phi_h\)). At that point, the stable inhomogeneous
structure \(\phi_1(y)\) and the trivial solution \(\phi_0(y)\)
\emph{exchange their relative stabilities}.

For completeness, in Fig.\ \ref{fig:2}b we plot
\(\mathcal{F}([\phi],k)\) for the same value of \(\phi_c/\phi_h\)
as in Fig.\ \ref{fig:2}a. For the chosen value of \(y_L\) it is
always \(\mathcal{F}^1(k)<0\), and correspondingly there is no
``stability exchange'' as a function of \(k\). Also, the initially
large \(\Delta\mathcal{F}^1(k)\) decreases with \(k\) as
\(\mathcal{F}^u(k)\) and \(\mathcal{F}^1(k)\) tend (for
\(k\to\infty\)) to the values corresponding to Dirichlet's b.c.\
\cite{IO}.
\subsection{Results for SSSR}\label{ssec:2.2}
By including an \emph{additive} spatiotemporal noise source
\(\xi(y,t)\) \cite{extend3b,nsp}, Eq.\ (\ref{eq:1}) becomes a
stochastic partial differential equation for the random field
\(\phi(y,t)\). The simplest assumptions about \(\xi(y,t)\) are that
it is Gaussian, with zero mean and a correlation function given by
\(\langle\xi(y,t)\xi(y',t')\rangle=2\gamma\,\delta(t-t')\delta(y-y')
\), where \(\gamma\) denotes the noise strength.

As discussed in \cite{extend2,extend2b,extend3a,extend3b}, known
results for activation processes in multidimensional systems
\cite{HG} allow us to estimate the activation rate using the
following Kramers'-like expression for the mean first-passage time
for the transitions between attractors
\[\langle\tau_i\rangle=\tau_0\,\exp
\left[\frac{\Delta\mathcal{F}^i(y_L)}{\gamma}\right],\] where
\(\Delta\mathcal{F}^i(y_L)=\mathcal{F}^u(y_L)-\mathcal{F}^i(y_L)\),
\(i=0,1\). The prefactor \(\tau_0\) is usually determined by the
curvature of \(\mathcal{F}[\phi]\) at its extrema. On one hand, it
is typically several orders of magnitude smaller than the average
time \(\langle\tau\rangle\), while on the other it does not change
significatively when varying the system's parameters around the
``bistable point'' \(y_L^*\), where \(\mathcal{F}([\phi_0],y_L^*)=
\mathcal{F}([\phi_1],y_L^*)\). Hence we may simplify the analysis
by assuming here that \(\tau_0\) is constant, and scale it out of
our results. The behavior of \(\langle\tau\rangle\) as a function
of \(k\) and \(\phi_c/\phi_h\) has been shown in
\cite{extend2,extend2b,IO}.

\begin{figure}
\centering
\resizebox{.4\columnwidth}{!}
{\includegraphics[bb= 0pt 0pt 301pt 231pt]{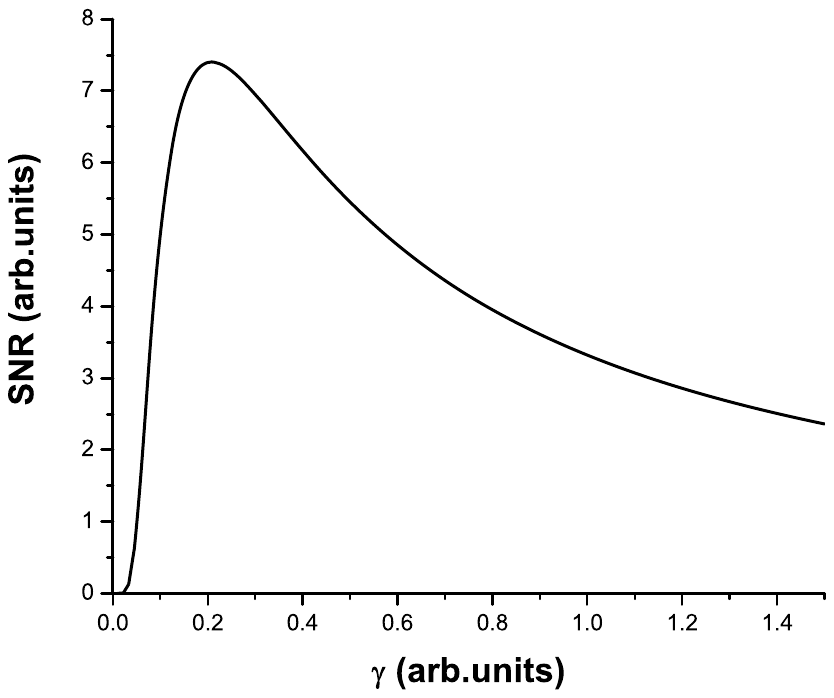}}
\qquad\qquad\resizebox{.4\columnwidth}{!}
{\includegraphics[bb= 135pt 111pt 571pt 674pt,
origin=rb,angle=-90]{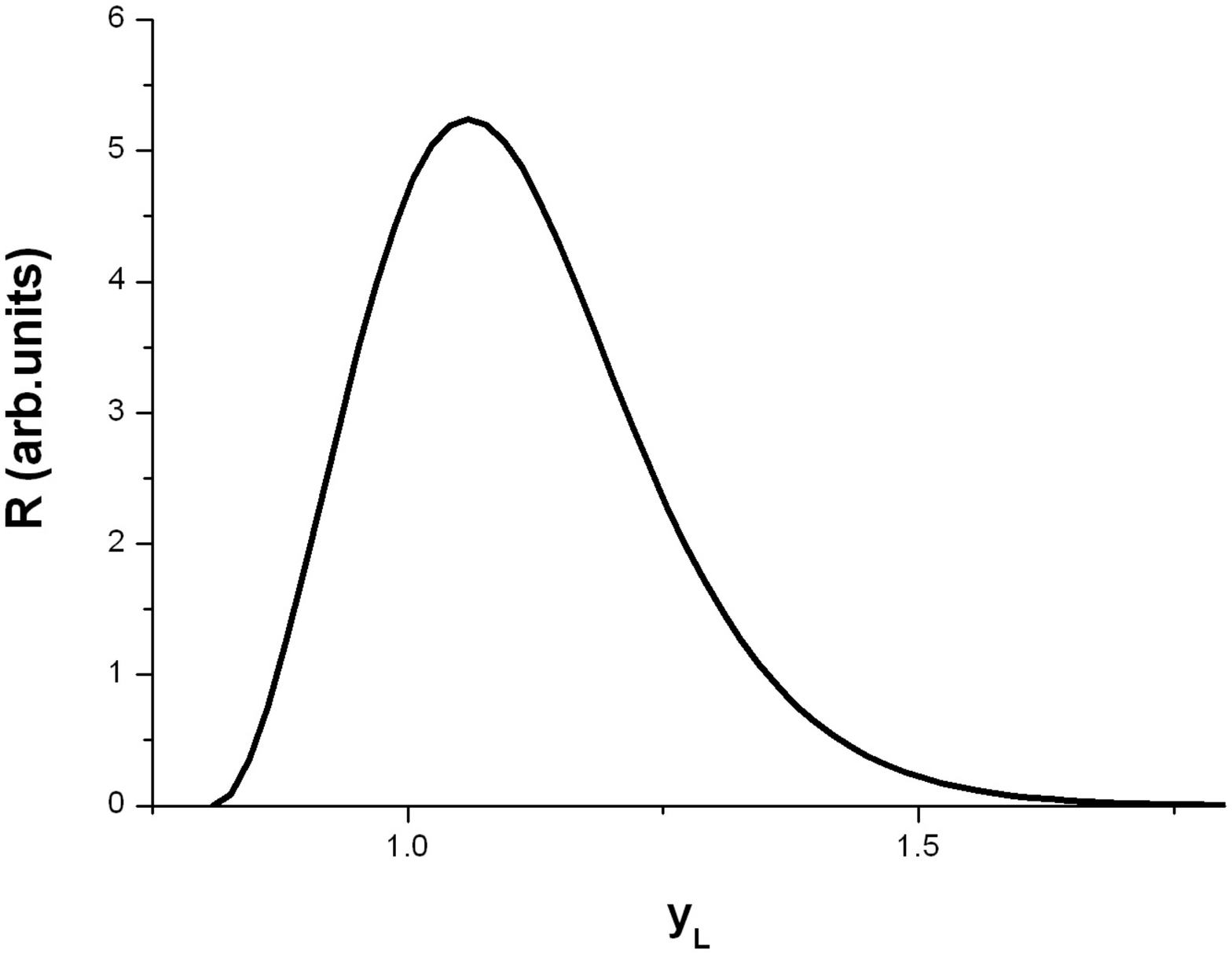}}
\caption{\textbf{Left}: SNR \emph{vs} the noise intensity
\(\gamma\), with system's size \(y_L=1.1\). \textbf{Right:} SNR
\emph{vs} \(y_L\), for \(\gamma=0.1\). The remaining parameters are
\(k=3.0\), \(D=1.0\), and \(\phi_c^0/\phi_h=0.193\).}\label{fig:3}
\end{figure}

As done in \cite{extend2}, we now assume that the system is
(adiabatically) subject to an external harmonic variation of the
parameter \(\phi_c\): \(\phi_c(t)=\phi_c^0+\delta\phi_c\cos(\omega
t)\) \cite{extend2b,extend3b}, and exploit the ``two-state
approximation'' \cite{RMP} as in \cite{extend2b,extend3a,extend3b}.
Such approximation reduces the whole dynamics on the bistable
potential landscape to one where the transitions occur only between
the states associated to the bottom of each well, hence the only
dynamical contents resides in the transition rates. Up to first
order in the amplitude \(\delta\phi_c\) (assumed to be small in
order that the periodic input be sub-threshold) the transition
rates \(W_i\) adopt the form
\begin{equation}\label{eq:5}
W_i\simeq\frac{1}{2}\left[\mu_i\mp\alpha_i\frac{\delta\phi_c}
{\gamma}\cos(\omega t)\right],
\end{equation}
where (at constant \(\phi_h\)) \(\mu_i\approx\exp[-\Delta\mathcal
{F}^i(\phi_c^0,y_L)/\gamma]\) and \(\alpha_i\approx\mu_i\,(\partial
\Delta\mathcal{F}^i/\partial\phi_c|_{\phi_c^0})\), \(i=0,1\). The
quantity inside parentheses can be obtained analytically using Eq.\
(\ref{eq:4}). These results allow to calculate the autocorrelation
function, the power spectrum density and finally the SNR, that we
indicate by \(R\). The detailed calculation can be found in the
appendix of \cite{extend3a}. Up to the relevant order (the second)
in the signal amplitude \(\delta\phi_c\), we obtain
\begin{equation}\label{eq:6}
R=\frac{\pi}{4\mu_0\mu_1}\,
\frac{(\alpha_0\mu_1+\alpha_1\mu_0)^2}{\mu_0+\mu_1}=
\frac{\pi}{4\gamma^2}\,\frac{\mu_0\mu_1}{\mu_0+\mu_1}\,\Phi,
\end{equation}
where we have used the form of the \(\alpha_i\) to reduce the
expression, and defined \(\Phi=[2\phi_h\,y_c(y_L)]^2\).
Figure \ref{fig:3} (left) is a plot of \(R\) as a function of the
noise intensity \(\gamma\) for a fixed system's length \(y_L\),
displaying the typical maximum that has become the fingerprint of
the SR phenomenon. In Fig.\ \ref{fig:3} (right), the roles of
\(\gamma\) and \(y_L\) are exchanged (\(R\) is plotted as a
function of \(y_L\) for fixed \(\gamma\)). Such a response is the
expected one for a system exhibiting SSSR. In both cases, the
values of \(k\) and \(\phi_c^0/\phi_h\) are kept fixed.

Within the NEP context and in this kind of systems, the phenomenon
arises due to the \emph{breakdown of the NEP's symmetry}. This
means that (as shown in Fig.\ \ref{fig:2}) when varying \(y_L\),
both attractors can exchange their relative stability. For
\(y_L=y_L^*\approx1\) both stable structures---the inhomogeneous
one \(\phi_1(y)\) and the trivial one \(\phi_0\)---have the
\emph{same} value for the NEP. For \(y_L<y_L^*\), \(\phi_1(y)\)
becomes less stable than \(\phi_0\) so transitions from
\(\phi_1(y)\) to \(\phi_0\) are more frequent (the barrier is
lower) than in the reverse direction, thus reducing the system's
response. When \(y_L\sim0.72\), \(\phi_1(y)\) and \(\phi_u(y)\)
coalesce and disappear, and the response is strictly zero (within
the linear response scheme implicit in the two-state
approximation). When \(y_L>y_L^*\), \(\phi_1(y)\) becomes more
stable than \(\phi_0\), making now transitions from \(\phi_0\) to
\(\phi_1(y)\) more frequent than in the reverse direction, and
reducing again the system's response. Clearly, the system's
response has a maximum when both attractors have the same stability
(\(y_L=y_L^*\)), and decays when departing from that situation.
Hence, for this system and within this framework, SSSR arises as a
particular case of the more general discussion done in
\cite{extend3a}. It should not come as a surprise to find an
analogy with the mechanism of double stochastic coherence described
in \cite{sscr}, where the NEP's symmetry is \emph{induced} by (an
additional, multiplicative) noise.

By comparing figures \ref{fig:2}a and \ref{fig:3} it becomes
apparent that the value of \(y_L\) at which the SNR has its maximum
\emph{differs slightly} from \(y_L^*\) (where the crossing between
\(\mathcal{F}^1\) and \(\mathcal{F}^0\) takes place). The origin of
this discrepancy is the following: whereas on qualitative grounds
we have argued that the maximum of the SNR should be related to the
potential being symmetric (both wells having the same ``energy'')
\cite{extend3a}, the exact condition is that the transition rates
between both wells be equal. In general, due to small differences
between the curvatures at the bottom of each well, those rates
become equal for values of \(y_L\) slightly different from the one at the symmetric case. Although by adopting here a constant value
of \(\tau_0\) we have assumed equal curvatures, there is still a
difference in the values of the \(\alpha_i\), since the \(\partial
\Delta\mathcal{F}^i/\partial\phi_c|_{\phi_c^0}\), \(i=0,1\) are
slightly different (a fact reflected in the dependence of \(\Phi\) on \(y_L\)).

\begin{figure}
\centering
\resizebox{.6\columnwidth}{!}{\includegraphics[bb= 100pt 0pt 500pt 842pt,angle=90]{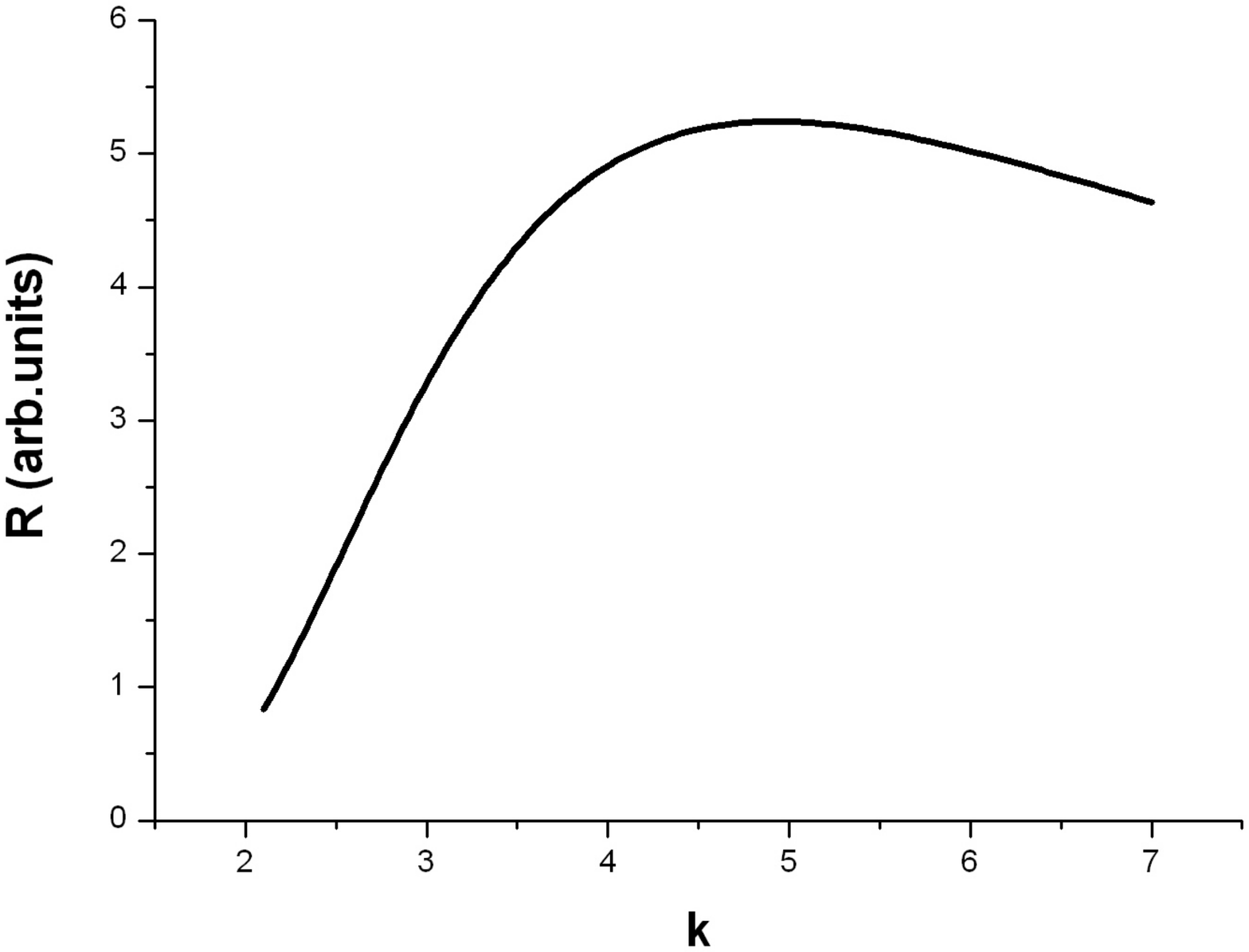}}
\caption{SNR \emph{vs} \(k\) for \(y_L=1.2\), \(\gamma=0.1\) and
\(\phi_c^0/\phi_h=0.193\).}\label{fig:4}
\end{figure}

Additional light can be shed on the phenomenon when viewed from a
different angle. Figure \ref{fig:4} is a plot of \(R\) as a
function of \(k\) at fixed values of \(\gamma\), \(y_L\) and
\(\phi_c^0/\phi_h\). It exhibits a broad resonance since for \(k\)
not too large (indicating a \emph{high reflectiveness} at the
boundary or a reduced exchange with the environment) \(R\)
increases with \(k\), whereas it slightly decreases for larger
\(k\) values (where the system's boundaries become absorbent). An
explanation of this behavior in terms of the NEP has been given in
\cite{SSSR6}: as already observed, the NEP's symmetry is broken for
this value of \(y_L\); moreover, whereas the lower branch in Fig.\
\ref{fig:2}b goes rapidly towards the value corresponding to
Dirichlet's b.c.\ the upper branch keeps increasing, thus degrading
the SNR. In any case, the fact that the resonance is broad
indicates the \emph{robustness} of the system's response with
regard to \(k\), a parameter that (together with \(\gamma\))
encodes the coupling with the environment.

We stress the fact that the NEP framework put forward in this
review allows to study SSSR between whole \emph{patterns}. The
explanation offered in \cite{SSSR4} to the phenomenon resorted to
a collective variable \(X\approx(1/N)\sum_{j=1}^Nx_j\), and to the
fact that the noise in the effective stochastic differential
equation for \(X\) scaled with size. In \cite{SSSR6} it was shown
that all the cases discussed in \cite{SSSR4} can be put within the
same NEP framework than the above studied scalar model. In fact,
the aforementioned almost linear \emph{increasing} dependence of
\(\Delta\mathcal{F}^1(y_L)\) on \(y_L\) can be interpreted as a
\emph{noise scaling} with size. There are however situations where
the NEP's symmetry is retained as the system's size is varied. We
may then speak of a \emph{genuinely noise-scaling} SSSR, in
contrast to the cases that could be called \emph{NEP symmetry
breaking} SSSR \cite{SSSR6}.
\section{Case of selective coupling}\label{sec:3}
In this section we analyze SR in two extended systems with
\emph{density-dependent} diffusive-like coupling: an extension of
the scalar RD model considered in Sec.\ \ref{sec:2}
\cite{extend3c}, and an array of FitzHugh--Nagumo \cite{FHN} units.
\subsection{Scalar model}\label{ssec:3.1}
Here we extend the one-component RD model discussed in Sec.\
\ref{sec:2} by letting the diffusive parameter \(D\) in Eq.\
(\ref{eq:1}) depend on the field \(\phi(x,t)\). As a matter
of fact, since in the ballast resistor \cite{FHN,SW} the thermal
conductivity is a function of the energy density, the resulting
equation for the temperature field includes a temperature-dependent
diffusion coefficient in a natural way. The form of the governing
equation is now
\begin{equation}\label{eq:7}
\partial_t\phi(x,t)=\partial_x\left[D(\phi)\partial_x\phi\right]+
f(\phi)+\xi(x,t),
\end{equation}
with \(\xi(x,t)\) and \(f(\phi)\) as in Sec.\ \ref{sec:2}.

As it was done for the reaction term, a simple choice (that retains
however the qualitative features of the system) is to consider the
following dependence of the diffusion term on the field variable
\[D(\phi)=D_0[1+h\,\Theta(\phi-\phi_c)].\]
For simplicity, here we choose the same threshold \(\phi_c\) for
the reaction term and the diffusion coefficient.

We assume the system to lie in a bounded domain \([-L,L]\), with
Dirichlet b.c.\ at both ends: \(\phi(\pm L,t)=0\). The form of the
patterns is analogous to what has been obtained in  Sec.\
\ref{sec:2}, the only difference being that in the present case
\(d\phi/dx|_{x_c}\) is \emph{discontinuous} and the area of the
``activated'' central zone depends on \(h\).

As before, the indicated patterns are extrema of the NEP: the
unstable pattern \(\phi_u(x)\) is a \emph{saddle-point} of this
functional, separating the \emph{attractors} \(\phi_0(x)\) and
\(\phi_s(x)\). For the case of a field-dependent diffusion
coefficient \(D(\phi(x,t))\) as described by Eq.\
(\ref{eq:7}), the NEP reads \cite{extend3c}
\[\mathcal{F}[\phi]=\int_{-L}^L\left\{-\int_0^\phi
D(\phi')f(\phi')\,d\phi'+\frac{1}{2}\left[D(\phi)
\frac{\partial\phi}{\partial x}\right]^2\,\right\}dx.\]
Given that \(\partial_t\phi=-[1/D(\phi)](\delta\mathcal{F}/
\delta\phi)\), one finds \(d\mathcal{F}/dt=-\int(\delta\mathcal{F}/
\delta\phi)^2\,dx\leq 0\), thus warranting the Lyapunov's
functional property.

Whereas in Sec.\ \ref{sec:2} we kept \(\phi_c^0\) constant and
varied \(y_L\), we now vary instead \(\phi_c^0\) at constant \(L\).
Similarly as before, both linearly stable states have the same
value of the NEP (i.e., they are equally stable) at some value
\(\phi_c^*\) of the threshold. The way \(\mathcal{F}[\phi]\)
depends on \(\phi_c^0\) resembles the dependence on \(y_L\) shown
in Sec.\ \ref{sec:2}, but now \(\Delta\mathcal{F}^s\) is an (almost
linearly) \emph{decreasing} function of \(\phi_c^0\) and both
inhomogeneous structures coalesce and disappear through a
\emph{subcritical} saddle--node bifurcation. As in the previous
case, we analyze only the neighborhood of \(\phi_c^*\). We shall
moreover consider only the neighborhood of \(h=0\), where the main
trends of the effect can be captured.

\begin{figure}
\includegraphics[bb= 14pt 14pt 581pt 529pt,width=5cm]{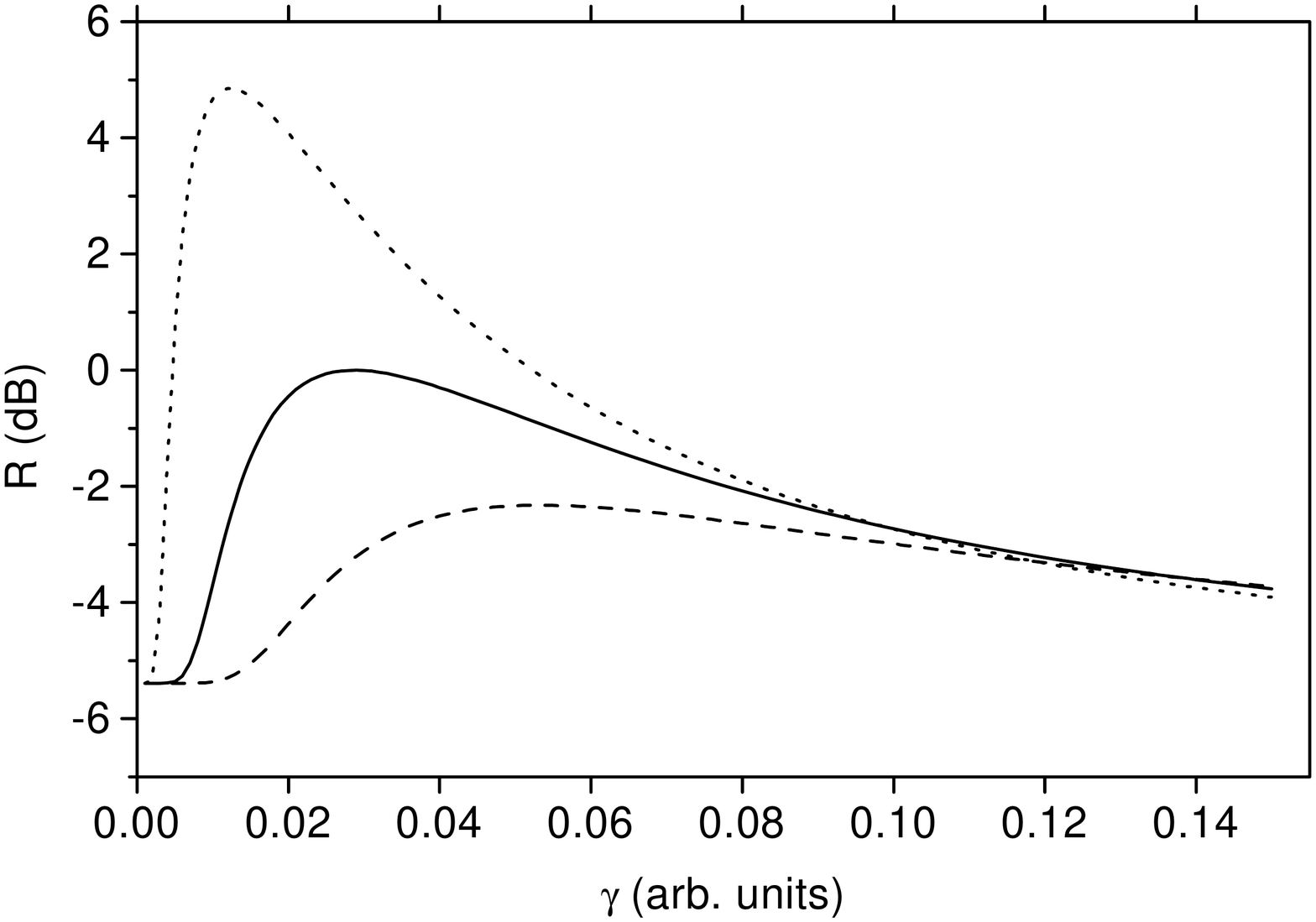}
\qquad\qquad\qquad
\includegraphics[bb= 14pt 14pt 581pt 529pt,width=5cm]{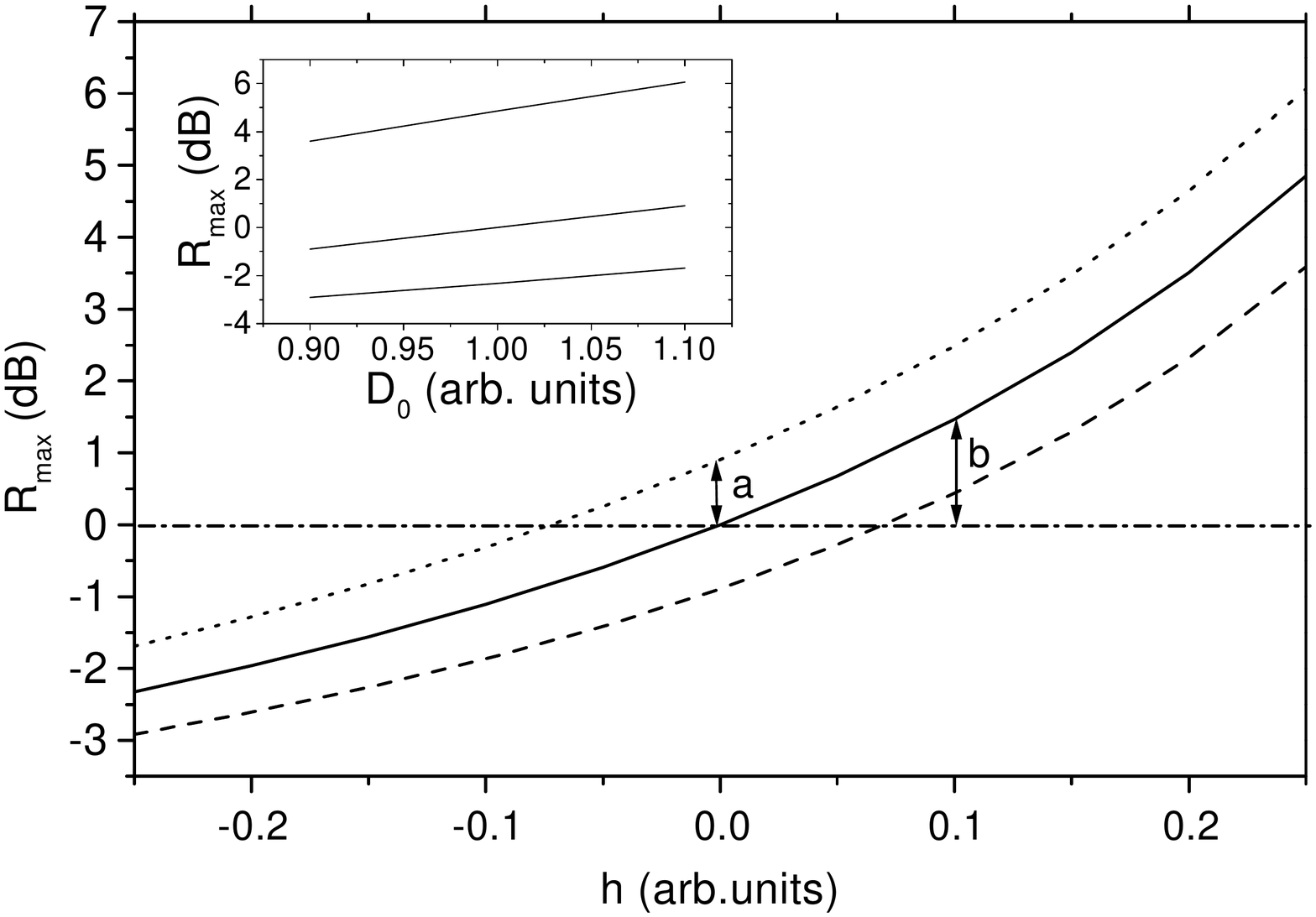}
\caption{\textbf{Left:} SNR \emph{vs} the noise intensity
\(\gamma\), for \(D_0=1.0\) and \(h=0.0\) (full line), \(-0.25\)
(dashed line) and 0.25 (dotted line). \textbf{Right:} the maximum
\(R_\mathrm{max}\) of the SNR curve as a function of the
selectiveness \(h\) of the coupling, for \(D_0=0.9\) (dashed line),
1.0 (full line) and 1.1 (dotted line). The arrows \textbf{a} and
\textbf{b} indicate the response gain due respectively to a
homogeneous increase of the coupling and to a selective one. The
larger gain in the second case is apparent. The inset shows the
dependence of \(R_\mathrm{max}\) on \(D_0\) for \(h=-0.25\)
(lower line), 0.0 and 0.25 (upper line). The remaining parameters
are \(L=1.0\), \(\delta\phi_c=0.01\) and \(\Omega=0.01\).}
\label{fig:5}
\end{figure}

Figure \ref{fig:5} (left) depicts the dependence of \(R\) on the
noise intensity \(\gamma\) for three values of \(h\), each curve
displaying the typical SR maximum. Figure \ref{fig:5} (right) is a
plot of the value \(R_\mathrm{max}\) of these maxima as a function
of \(h\). The dramatic increase of \(R_\mathrm{max}\) (several dB
for a \emph{small} positive variation of \(h\)) is apparent, and
shows the strong effect that the selective coupling (or
field-dependent diffusivity) has on the system's response.

It must be noted that the only two approximations made in order to
derive our results---namely the Kramers-like expression
and the two-level approximation used for the evaluation of the
correlation function \cite{extend3c}---break down for
large positive values of \(h\) because for increasing selectivity
the curves of \(\mathcal{F}[\phi]\) \emph{vs} \(\phi_c^0\) shift
towards the left \cite{extend3c}, which in turn means that the
barrier separating the attractors at \(\phi_c^*\) tends to zero.
This effect is basically the same as the one discussed in Refs.\
\cite{extend2b,extend3a} in connection with global diffusivity
\(D_0\). It is also worth noting that except for the two
aforementioned approximations, all the previous results (e.g.\ the
profiles of the stationary patterns and the corresponding values of
the nonequilibrium potential) are analytically exact.
\subsection{FitzHugh--Nagumo model}\label{ssec:3.2}
Here we study an array of FitzHugh--Nagumo \cite{FHN} units, with a
density-dependent (diffusive-like) coupling. The NEP for this
system was found within the excitable regime and for particular
values of the coupling strength \cite{IO2}. In the general case,
however, the form of the NEP has not been found yet. Hence, we have
resorted to a study based on numerical simulations, analyzing the
influence of different parameters on the system's response.
Nevertheless, the idea of the existence of such a NEP has always
underlied this study. The results show that the enhancement of the
SNR found for the scalar system \cite{extend3c} is \emph{robust},
and that the indicated non-homogeneous coupling could clearly
contribute to enhance the SR phenomenon in more general situations.

We consider a simplified version of the FitzHugh--Nagumo model
\cite{extend3c,IO,FHN}, which has been useful for gaining
qualitative insight into the excitable and oscillatory dynamics in
neural and chemical systems. It consist of two variables:
\begin{itemize}
\item a (fast) \emph{activator} field \(u\), that in the case of
neural systems represents the voltage variable, while in chemical
systems represents the concentration of an autocatalytic species.
\item an \emph{inhibitor} field \(v\), associated (within a neural
context) to the concentration of potassium ions in the medium, and
that in a general chemical reaction inhibits the generation of the
\(u\) species.
\end{itemize}
Instead of considering the usual cubic-like nonlinear form, we use
a piecewise linear version
\begin{eqnarray}
\epsilon \,\frac{\partial u(x,t)}{\partial t}&=&
\frac{\partial}{\partial x}\left[D_u(u)\,\frac{\partial u}
{\partial x}\right]+f(u)-v+\xi(x,t)\label{eq:8}\\
\frac{\partial v(x,t)}{\partial t}&=&\frac{\partial}{\partial x}
\left[D_v(v)\,\frac{\partial u}{\partial x}\right]+\beta\,u-
\alpha\,v,\label{eq:9}
\end{eqnarray}
where \(f(u)=-u+\Theta(u-\phi_c)\), and \(\xi(x,t)\) is a
\(\delta\)--correlated white Gaussian noise, as before. \(\gamma\)
indicates the noise intensity and \(\phi_c\) is the
``discontinuity'' point, at which the piecewise linearized function
\(f(u)\) presents a jump. The parameter
\(\epsilon=\tau_u/\tau_v\ll1\) indicates the timescale ratio
between the (fast) activator and the inhibitor. Together with
\(\alpha\) and \(\beta\), it is chosen to correspond to the
\emph{excitable} regime. We consider Dirichlet b.c.\ at
\(x=\pm L\). Although the results are qualitatively the same as
those that could appear considering the usual FitzHugh--Nagumo
equations, this simplified version allows us to compare directly
with the previous analytical results for this system
\cite{extend3c,extend3b}.

As in \cite{extend3c}, we assume that the diffusion coefficient
\(D_u(u)\) is not constant, but depends on the field \(u\)
according to \(D_u(u)=\,D_u^0\left[1+h\,\Theta(u-\phi_c)\right]\).
This form implies that the value of \(D_u(u)\) depends
``selectively'' on whether the field \(u\) fulfills \(u>\phi_c\) or
\(u<\phi_c\). \(D_u^0\) is the value of the diffusion constant
without such ``selective'' term, and \(h\) indicates the size of
the difference between the diffusion constants in both regions [if
\(h=0\) then \(D_u(u)=D_u^0\) constant]. \(D_v(v)\) is the
diffusion for the inhibitor \(v\), that we assume to be
homogeneously constant.

This system is known to exhibit two stable stationary patterns. One
of them is \(u(x)=0\), \(v(x)=0\), while the other is one with
nonzero values. Further, we consider, as before, that an external,
periodic, signal enters into the system through the value of the
threshold \(\phi_c\), \(\phi_c(t)=\phi_c^0+\delta\phi\,\cos(\omega
t)\), where \(\omega\) is the signal frequency, and \(\delta\phi\)
its intensity.

All the results were obtained through numerical simulations of the
system. The continuous version of the system indicated by Eqs.\
(\ref{eq:8}), (\ref{eq:9}) was transformed into a
second-order spatially discrete one
\begin{eqnarray*}
\dot{u_i}&=&D_{u_i}\Delta u_i+(D_{u_{i+1}}-D_{u_{i-1}})
(u_{i+1}+u_{i-1})+f(u_i)-v_i+\xi_i(t),\\
\dot{v_i}&=&D_v\Delta v_i+\beta\,u_i-\alpha\,v_i,
\end{eqnarray*}
with \(\Delta\phi_i\equiv(\phi_{i-1}+\phi_{i+1}-2\phi_i)\). The
extensive numerical simulations performed for a set of equations
were done exploiting Heun's algorithm \cite{nsp}.

In this spatially-extended system there are different ways of
measuring the overall system's response to the external signal. In
particular, we show the evaluated output SNR in two different ways
(the units being given in dB):
\begin{itemize}
\item SNR for the middle element of the chain evaluated over the
dynamical evolution of \(u_{N/2}\), that we call SNR\(_2\)
(however, having Dirichlet b.c., the local response depends on the
distance to the boundaries).
\item In order to measure the overall response of the system to the
external signal, we computed the SNR as follows: We digitized the
system's dynamics to a dichotomic process \(s(t)\): At time \(t\)
the system has an associated value of \(s(t)=1\,(0)\) if the
Hilbert distance to pattern \(1\,(0)\) is lower than to the other
pattern. Stated in mathematical terms, we computed the distance
\(\mathcal{D}_2[\cdot,\cdot]\) defined by
\[\mathcal{D}_2[f,g]=\left\{\int_{-L}^Ldx\,\left[f(x)-g(x)\right]^2
\right\}^{1/2}\] in \(\mathcal{L}^2([-L,L])\), the Hilbert space of
the real-valued functions in that interval. At time \(t\), a
digitized process is computed by means of
\[s(t)=\left\{\begin{array}{l}
1\:\hbox{ if }\:\mathcal{D}_2\left[P^u_1(x),u(x,t)\right]
<\mathcal{D}_2\left[P^u_0(x),u(x,t)\right]\\
0\:\hbox{ if }\:\mathcal{D}_2\left[P^u_1(x),u(x,t)\right]
\geq\mathcal{D}_2\left[ P^u_0(x),u(x,t)\right]\end{array}\right.,\]
We call this ``global-like'' measure SNR\(_p\).
\end{itemize}
The parameters kept fixed have been summarized in Table \ref{tb:1}.
The simulation was repeated 250 times for each parameter set, and
the SNR was computed by recourse of the average power spectral
density.

\begin{figure}
\centering
\resizebox{.4\columnwidth}{!}
{\includegraphics[width=4.5cm]{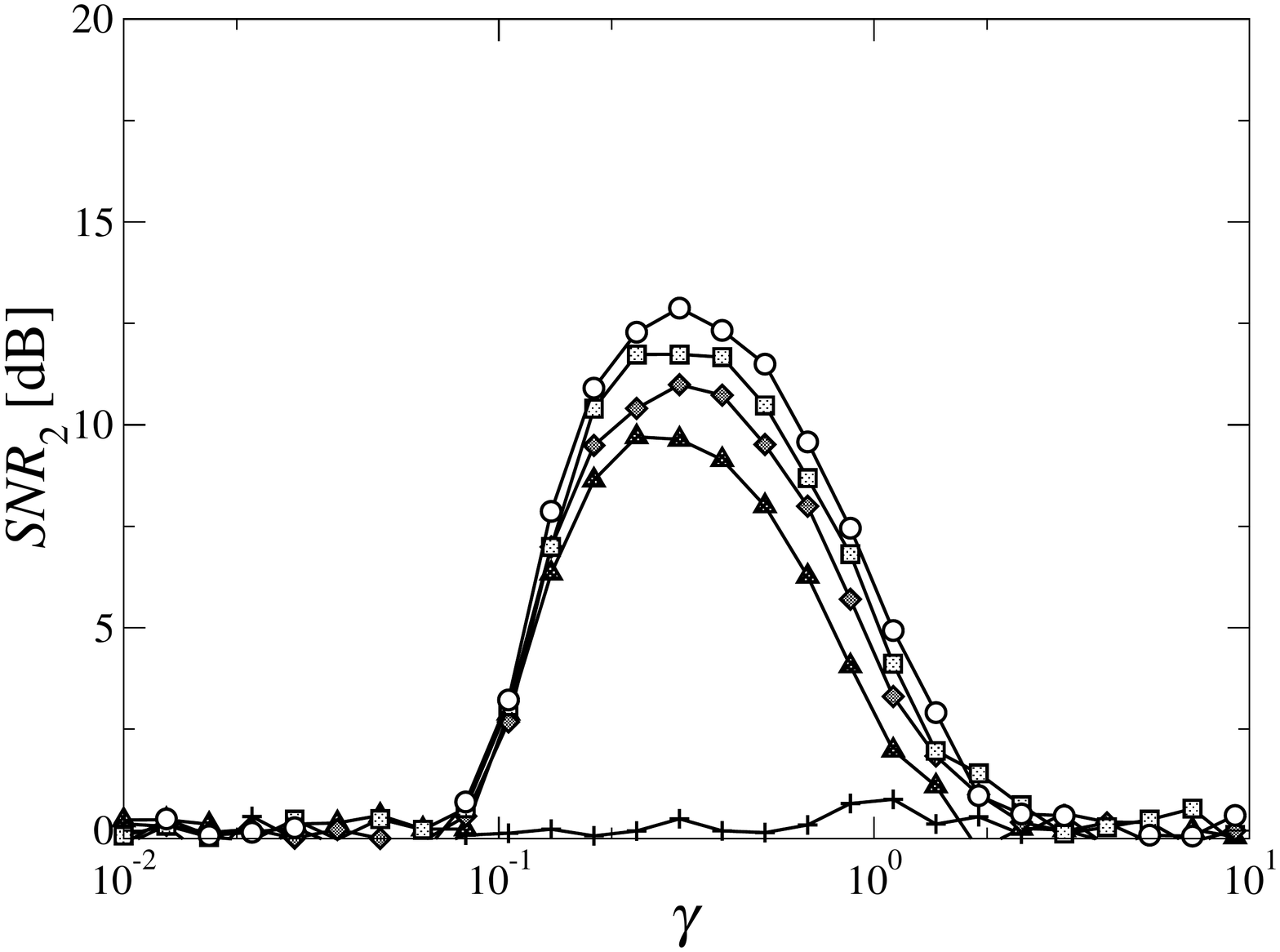}}
\resizebox{.4\columnwidth}{!}
{\includegraphics[width=4.5cm]{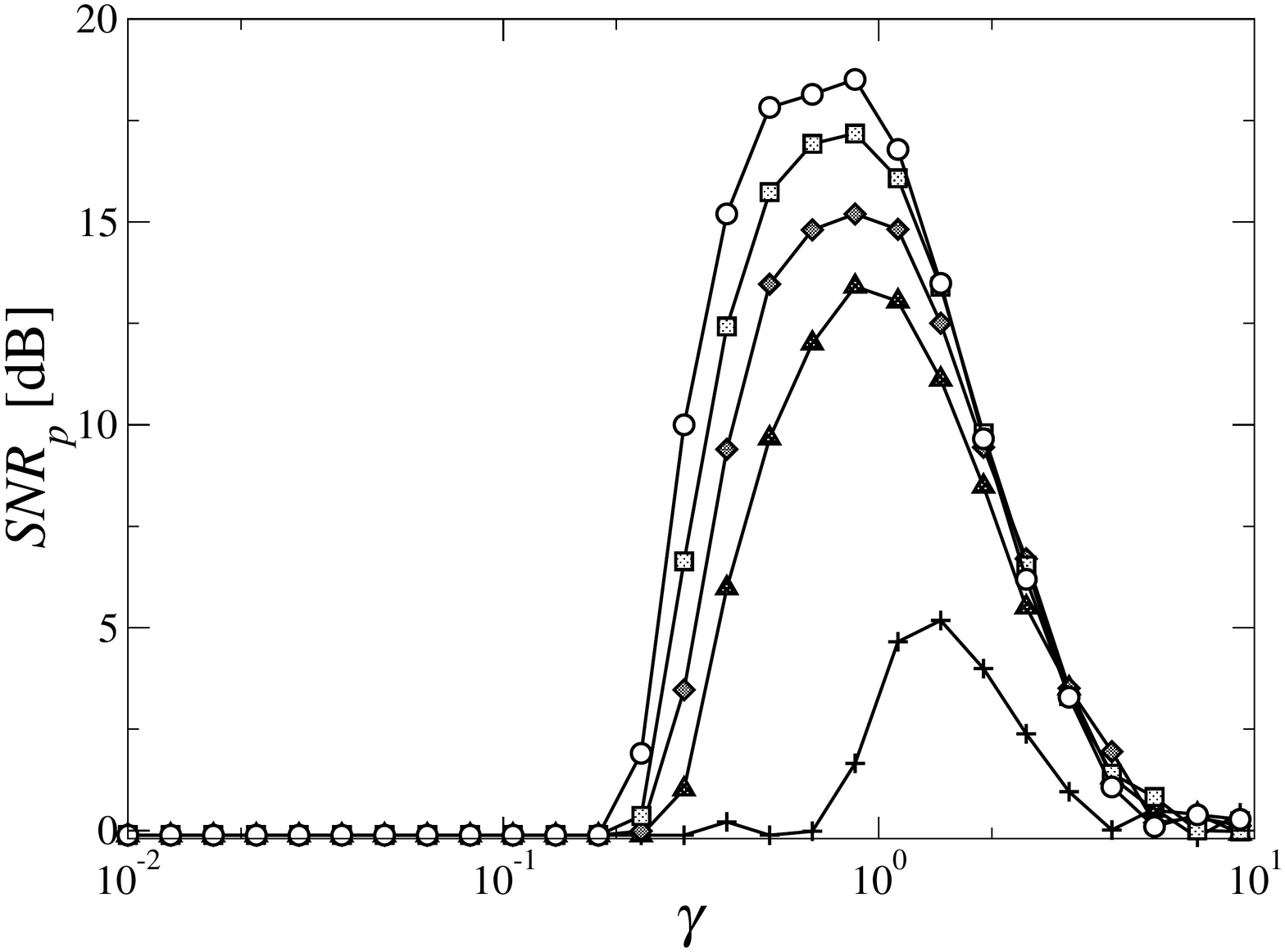}}
\caption{SNR\(_2\) and SNR\(_p\) \emph{vs} the noise intensity
\(\gamma\), for \(D_u^0=0.3\) and \(h=-2.0\) (+), \(-1.0\)
(\(\triangle\)), 0.0 (\(\diamond\)), 1.0 (\(\square\)), and 2.0
(\(\bigcirc\)). Both measures reveal a systematic enhancement of
the SNR as \(h\) increases.}\label{fig:6}
\end{figure}

Figure \ref{fig:6} depicts the results for the different
SNR measures we have previously defined, as functions of the noise
intensity \(\gamma\). For both measures it is apparent that there
is an enhancement of the response for \(h>0\), when compared with
the \(h=0\) case, while for \(h<0\) the response is smaller.
\begin{table}
\centering
\begin{tabular}{|ccccccc|cc|}\hline
\multicolumn{7}{|c|}{Model}&\multicolumn{2}{c|}{Numerical}\\\hline
\(\alpha\)&\(\beta\)&\(\epsilon\)&\(\phi_c^0\)&\(\delta\phi\)&
\(\omega\)&\(D_v\)&\(\Delta t\)&\(N\)\\
0.3&0.4&0.03&0.52&0.4&\(5\pi/8\)&1.0&\(10^{-3}\)&51\\
\hline\end{tabular} \caption{Fixed parameters for the
FitzHugh--Nagumo model with \(D_u(u)\)}\label{tb:1}
\end{table}

\begin{figure}
\centering
\resizebox{.4\columnwidth}{!}
{\includegraphics[width=4.5cm]{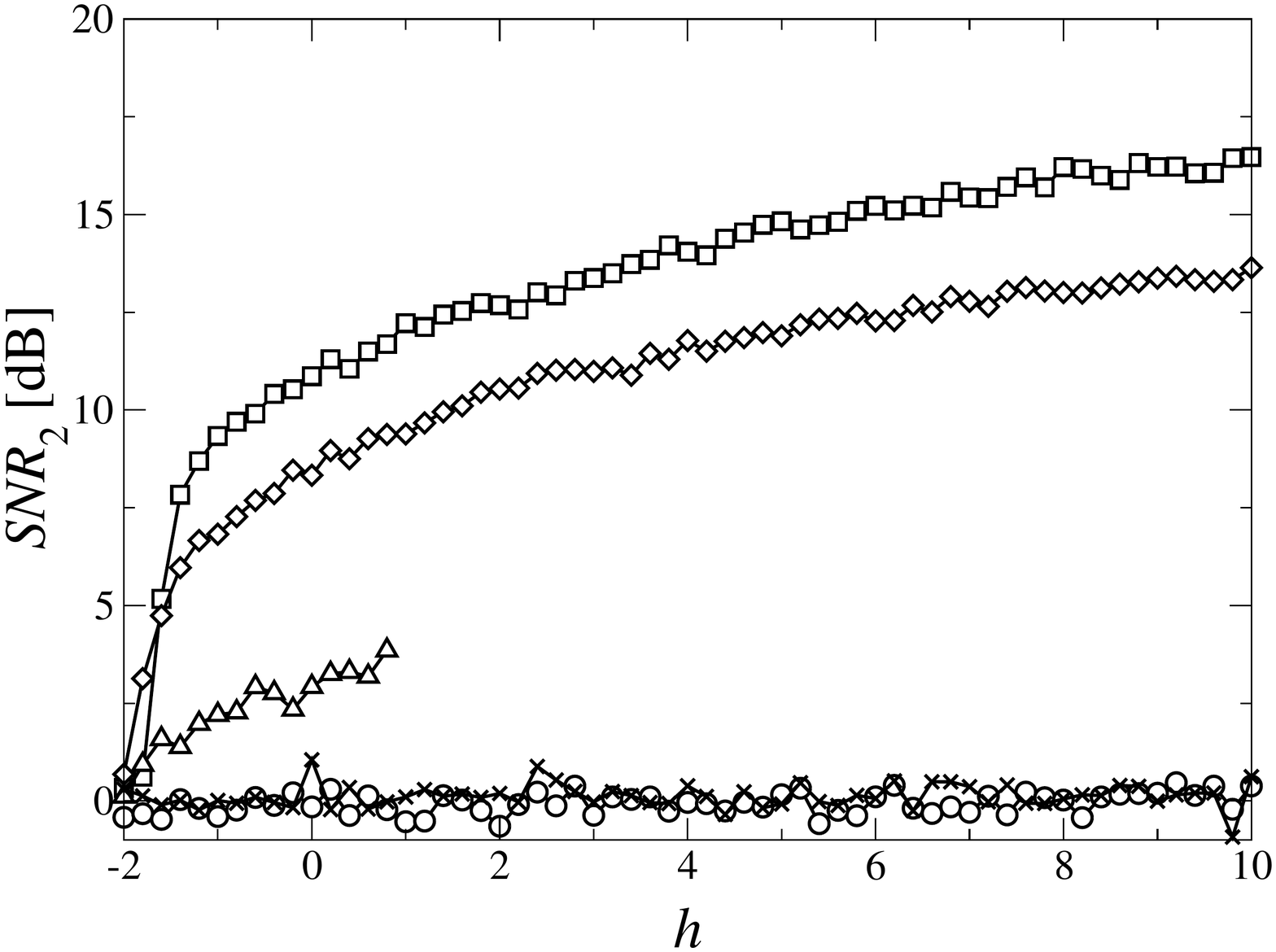}}
\resizebox{.4\columnwidth}{!}
{\includegraphics[width=4.5cm]{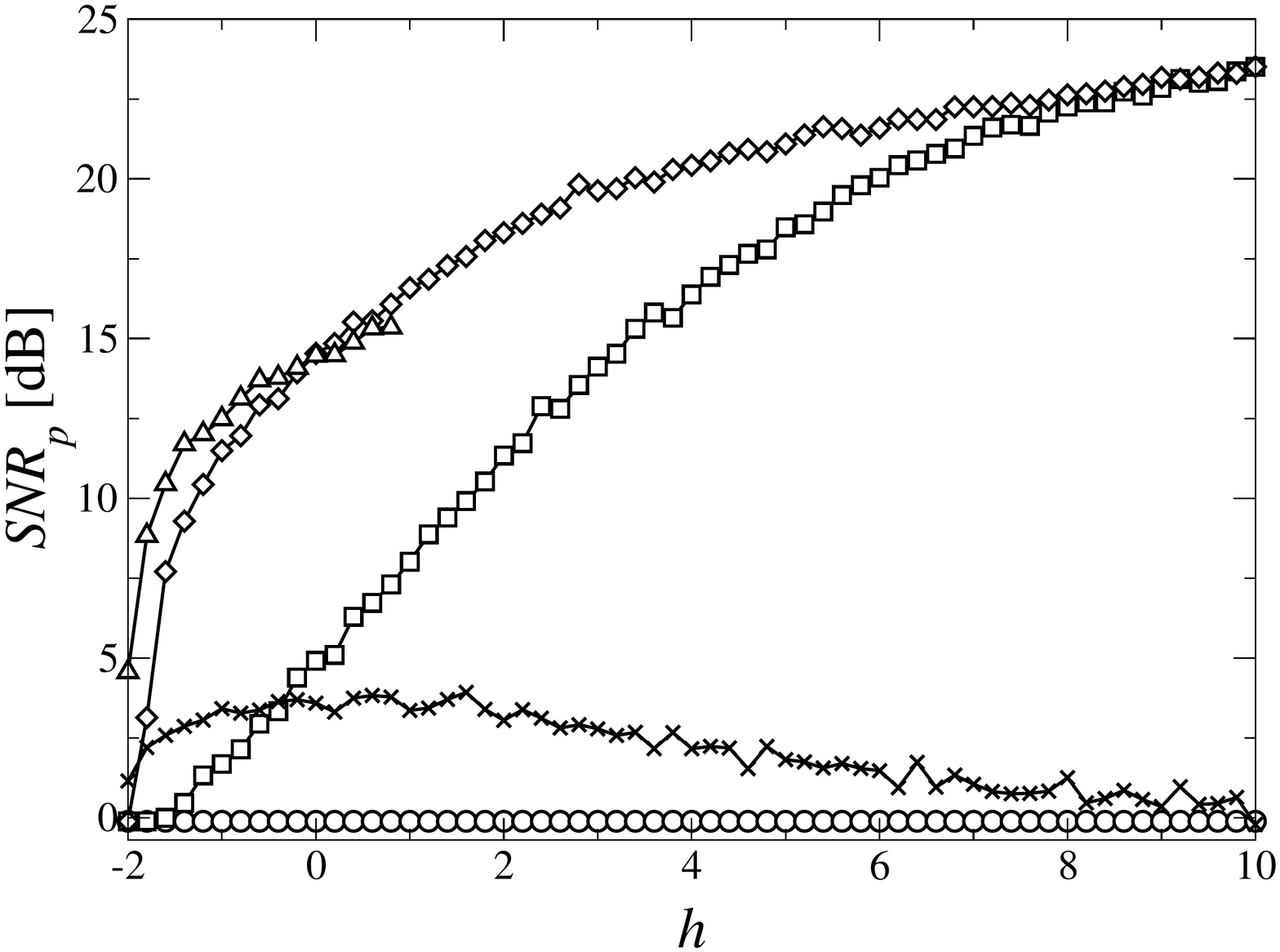}}
\caption{SNR\(_2\) and SNR\(_p\) \emph{vs} the selectiveness of
coupling \(h\), for \(D_u^0=0.3\) and \(\gamma=0.032\)
(\(\bigcirc\)), 0.32 (\(\square\)), 0.6 (\(\diamond\)), 1.2
(\(\triangle\)), and 3.2 (\(\times\)).}\label{fig:7}
\end{figure}

In Fig.\ \ref{fig:7} we show again the response's measures, but
now as functions of \(h\). We have plotted the maximum of each SNR
curve for \(D_u^0=0.3\), and \(\gamma=0.01\), 0.1, and 0.3. It is
clear that there exists an optimal value of \(\gamma\) for which
the response is largest. The rapid fall in the response for \(h<0\)
is also apparent.
\begin{figure}
\centering
\resizebox{.4\columnwidth}{!}
{\includegraphics[width=4.5cm]{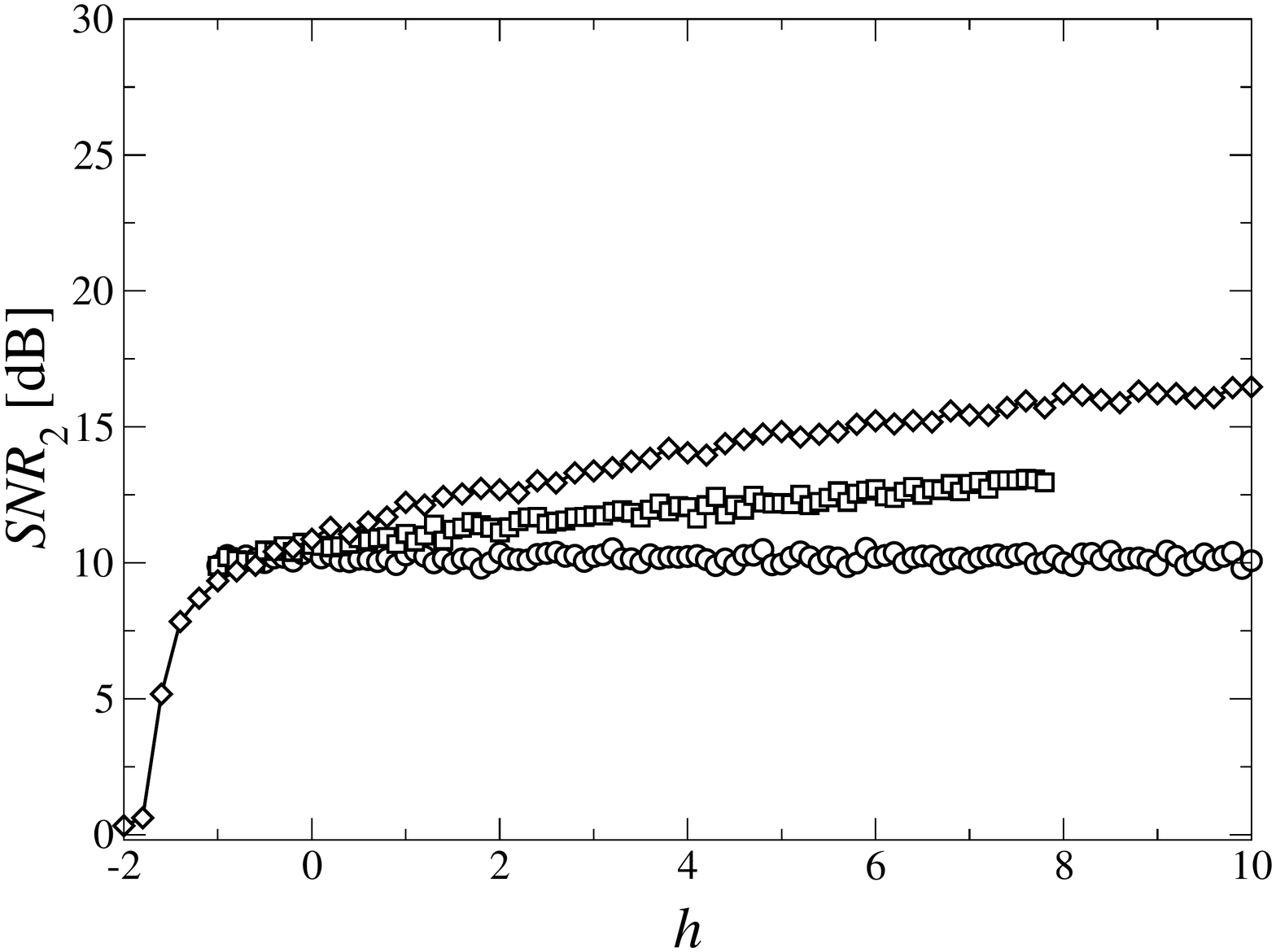}}
\resizebox{.4\columnwidth}{!}
{\includegraphics[width=4.5cm]{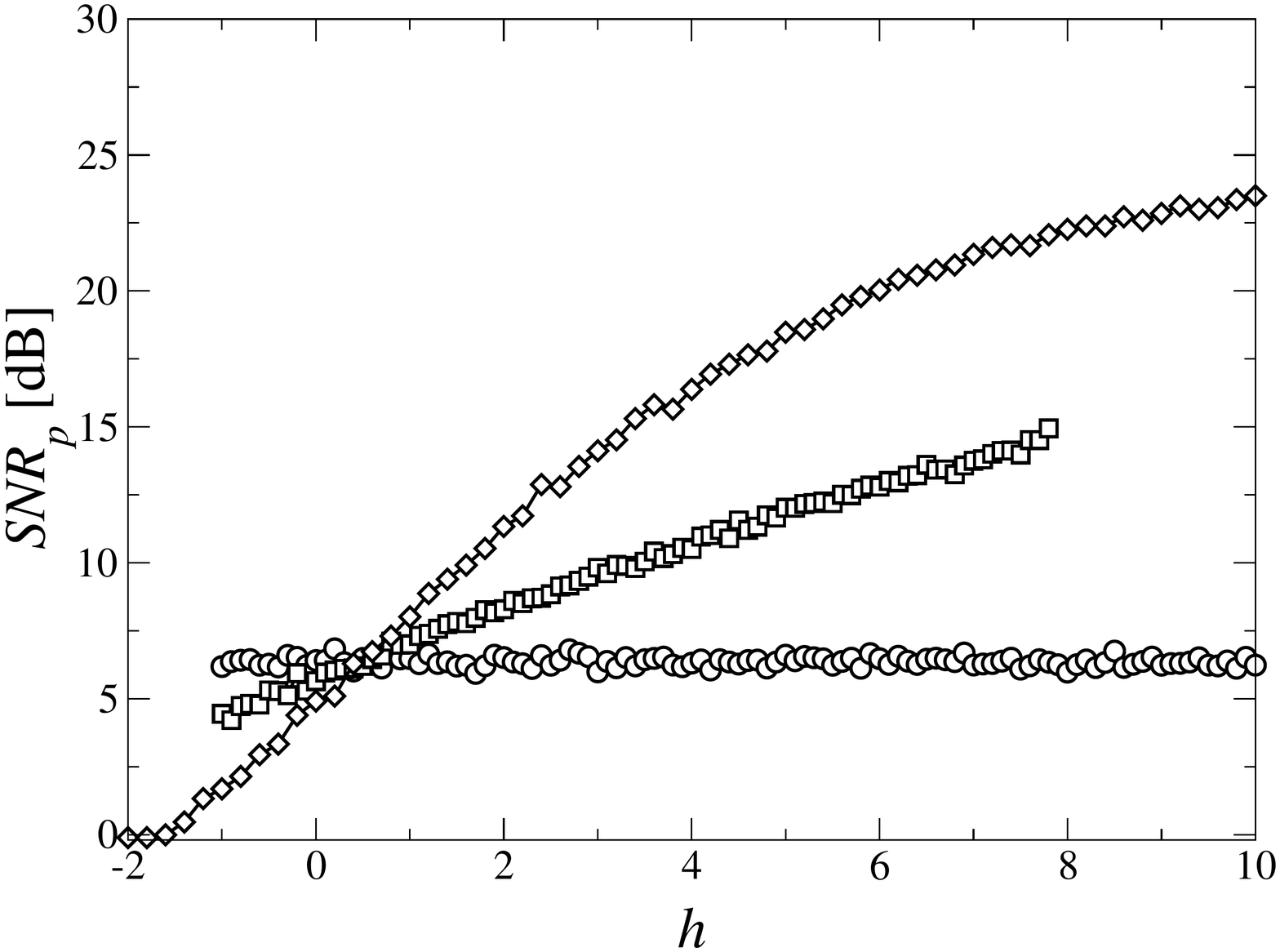}}
\caption{SNR\(_2\) and SNR\(_p\) \emph{vs} \(h\) for
\(\gamma=0.032\) and \(D_u^0=0.0\) (\(\bigcirc\)), 0.1
(\(\square\)), and 0.3 (\(\diamond\)).}\label{fig:8}
\end{figure}

In Fig.\ \ref{fig:8} we show the dependance of SNR on \(h\),
for different values of the diffusion which depends on the
activator density \(D_u^0\). It is apparent that the response
becomes larger when the value of \(D_u^0\) is larger. However, as
was discussed in \cite{extend2,extend3b}, it is clear that for
still larger values of \(D_u^0\), the symmetry of the underlying
potential (that is the relative stability between the attractors)
is broken and the response finally falls down.

The previous figures clearly show that the response to the external
signal grows with the ``selectiveness'' of the coupling, showing
the robustness of the phenomenon presented in
 \cite{extend3c,extend3b}.
\section{Nonlocal Interaction}\label{sec:4}
Let us consider again a system like the one described by Eqs.\
(\ref{eq:8}), (\ref{eq:9}), but now assume that \(D_u\) and
\(D_v\) are constant. In Ref.\ \cite{IO1} it was assumed that the
inhibitor-like field has a \emph{diffusive} transport behavior, and
is \emph{fast} enough that can be adiabatically eliminated, thus
yielding an effective scalar RD equation with a \emph{nonlocal}
term, characterized by a diffusive kernel \(G(x,x')\). After
briefly reviewing the derivation of the NEP for that situation, we
shall assume in this section that the transport mechanism of the
adiabatically eliminated inhibitor-like field is of
\emph{nondiffusive} character, thus yielding a kernel \(H(x,x')\)
that is more localized in space, and with a controllable
interaction range.

Following Ref.\ \cite{IO1}, let the system be defined by
\begin{eqnarray}
\partial_tu(x,t)&=&D_u\partial^2_xu(x,t)-u(x,t)+\Theta[u(x,t)-a]-
v(x,t),\nonumber\\
\epsilon^{-1}\,\partial_tv(x,t)&=&D_v\partial^2_xv(x,t)+\beta
u(x,t)-\alpha v(x,t),\label{eq:10}
\end{eqnarray}
where \(\epsilon\) was defined after Eqs.\ (\ref{eq:8}),
(\ref{eq:9}), and let it be confined to the domain \([-L,L]\),
with Dirichlet b.c.\ at both ends: \(u(\pm L,t)=v(\pm L,t)=0\).
Contrarily to the standard hypothesis, we now assume that \emph{the
inhibitor is much faster than the activator} (i.e.\
\(\tau_v\ll\tau_u\)). In the limit \(\epsilon\to\infty\), we can
rewrite Eq.\ (\ref{eq:10}) as
\begin{eqnarray*}
\partial_tu(x,t)&=&D_u\partial^2_xu(x,t)-u(x,t)+\Theta[u(x,t)-a]-
v(x,t),\\
0&=&D_v\partial^2_xv(x,t)+\beta u(x,t)-\alpha v(x,t).
\end{eqnarray*}
In the last pair of equations we can eliminate the inhibitor (which
is now \emph{slaved} to the activator) by solving the second
equation using the Green's function method
\begin{eqnarray*}
[-D_v\partial^2_x+\alpha]\,G(x,x')&=&\delta(x-x'),\\
v(x)&=&\beta\int dx'G(x,x')u(x'),
\end{eqnarray*}
where the Green's function \(G(x,x')\) is given by
\[G(x,x')=\frac{1}{D_vk}\left\{\begin{array}{lll}
\left[\sinh k(L-x')/\sinh 2kL\right]\sinh k(L+x)&&x<x',\\
\left[\sinh k(L+x')/\sinh 2kL\right]\sinh k(L-x)&&x>x',\\
\end{array}\right.\]
with \(k=(\alpha/D_v)^{1/2}\). This slaving procedure reduces our
system to a \emph{nonlocal} equation for the activator only, that
has the form
\begin{equation}\label{eq:11}
\frac{\partial u(x,t)}{\partial t}=D_u\,\frac{\partial^2u(x,t)}
{\partial x^2}+f(u)-\beta\int_{-L}^LG(x,x')\,u(x')\,dx'.
\end{equation}
From this equation, and taking into account the symmetry of the
Green's function \(G(x,x')\), we can obtain the Lyapunov functional
for this system, which has the form
\begin{equation}\label{eq:12}
\mathcal{F}[u]=\int_{-L}^Ldx\left[\frac{D_u}{2}
\left(\frac{\partial u}{\partial x}\right)^2-\int^uf(w)\,dw+
\frac{\beta}{2}\int_{-L}^Ldx'\,G(x,x')\,u(x')\,u(x)\right].
\end{equation}
This spatial nonlocal term in the NEP takes into account the
repulsion between activated zones. When two activated zones come
near each other, the exponential tails of the inhibitor
concentration overlap, increasing its concentration between both
activated zones and creating an effective repulsion between them.
Hence the Green's function plays the role of an exponential
screening between the activated zones. In Ref. \cite{extend2b} the
knowledge of such NEP was exploited to study SR on the system
indicated by Eq.\ (\ref{eq:11}).

The starting point of our present analysis will be the effective,
nonlocal and stochastic RD equation for the real (activator-like)
field \(\phi(x,t)\), analogous to Eq.\ (\ref{eq:11}), defined in
the one dimensional domain \(x\in[-L,L]\) by
\[\frac{\partial\phi}{\partial t}=D\,\frac{\partial^2\phi}{\partial
x^2}+f(\phi)-\beta\int_{-L}^LH(x,x')\,\phi(x')\,dx'+\xi(x,t),\]
where the diffusivity \(D\) is constant, and we assume a
\emph{cubic} nonlinear term \(f(\phi)=\phi\,(\phi-b)\,(2-\phi)\).
Here \(\xi(x,t)\) is an additive Gaussian white noise, as in the
previous cases. As before, the system is subject to Dirichlet b.c.\
\(\phi(\pm L,t)=0\).

Similarly to Eq.\ (\ref{eq:11}), this system could be written in
a variational form, with the functional \(\mathcal{F}[\phi]\) given
by Eq.\ (\ref{eq:12}). As anticipated, here we consider a
\emph{nondiffusive} kernel, with a controllable interaction range.
In order to keep our analysis simple we propose the following form
\begin{equation}\label{eq:13}
H(x,x')=\left\{\begin{array}{cl}
1/2;&|x-x'|\leq l\\
0;&|x-x'|>l,
\end{array}\right.
\end{equation}
which allows the analysis by just varying the interaction range
\(2l\).

The new effective RD equation contains local and nonlocal couplings
(corresponding to the diffusive and the nonlocal contribution,
respectively). The last one contains the nonlocal kernel given by
Eq.\ (\ref{eq:13}), with a variable range \(2l\), that
corresponds to the interaction of the field at points
\(x'\in[x-l,x+l]\). However, such points will contribute if and
only if they are inside the domain \([-L,L]\). We are now in
position to study the role played by the nonlocal kernel
(particularly by its range \(2l\)) on the SR phenomenon.

As before, the SR between stationary solutions was investigated in
terms of the two-state approach (all the details about the
procedure and the evaluation of the SNR can be found in Refs.\
\cite{extend3a,extend3b}). As usual, we subject our system to a
weak external signal \(b=b_0+A(t)=b_0+\Delta b\,\cos(\omega_st)\),
rocking the NEP \cite{extend3b}. In order to have a subthreshold
signal, the amplitude \(\Delta b\) should satisfy \(\Delta b\ll
b\). We have chosen \(b_0\) as the value of \(b\) at which
\(\mathcal{F}[\phi_s]=\mathcal{F}[\phi=0]=0\) when \(\beta=0\).

Up to first order in the small amplitude \(\Delta b\), the
transition rates \(W_i\) and the functions \(\alpha_i\) have the
form indicated in Eqs.\ (\ref{eq:5}), (\ref{eq:6}). But now
\(\Phi\), that depends on the inhomogeneous attractor \(\phi_s\),
has the form \[\Phi=\left[\int_{-L}^L\phi_s^2
\left(1-\frac{\phi_s}{3}\right)dx\right]^2.\]

After fixing the length of the system \(L\) and the kernel range
\(2l\) we can use the above indicated expressions to find the SNR,
that shows the usual bell-shaped form of stochastic resonance as a
function of the noise intensity.

\begin{figure}[!ht]
\centering
\includegraphics[bb= 54pt 360pt 558pt 720pt,width=6cm]{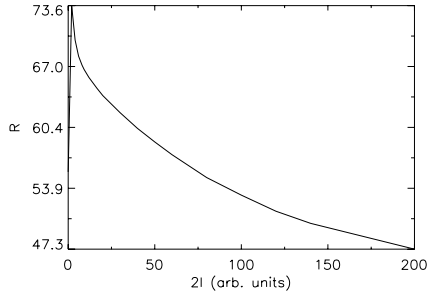}
\caption{SNR as a function of \(2l\) for \(2L\) fixed. The
parameters are \(D=0.6\), \(2L=6.25\), \(\beta=0.02\) and
\(b_0=0.719123\).}\label{fig:9}
\end{figure}

In Fig.\ \ref{fig:9} we show the dependence of the SNR on the
kernel range \(2l\) for a fixed value of \(2L\). There is a
nonmonotonic behavior in the system's response against variation of
\(2l\), that can be explained by the following facts:
\begin{itemize}
\item On one hand, the transition rates are decreasing functions of
the range \(2l\) for fixed \(2L\). Therefore, the ratio
\(\mu_1\mu_2/(\mu_1+\mu_2)\) in the expression for the SNR also
reflects this behavior.
\item The other factor in this expression has a maximum for a
kernel range that corresponds to ``first neighbor'' sites, namely
\(x=x'\pm l\).
\end{itemize}
The maximum in the system's response as a function of the kernel
range is due to the interplay between these two factors. From this
analysis of the comparative weights of the local (diffusive) and
nonlocal terms contributing to the SR response, it is apparent that
the range of such nonlocal kernel has an optimum value yielding a
maximum for the SNR \cite{bvhhswnew}.
\section{Conclusions}\label{sec:5}
We have discussed three different aspects of the phenomenon of
stochastic resonance in reaction--diffusion systems, within the
nonequilibrium potential's framework. In first place we have
discussed \emph{system-size SR} in a scalar model. Even though we
have not shown the details here, it has been also possible to also
study other cases \cite{SSSR6}. In particular, a model of globally
coupled nonlinear oscillators discussed in \cite{SSSR4}, showing
that it can also be described within the NEP framework, with SSSR
arising through an ``effective'' scaling of the noise intensity
with the system's size.

In second place we presented a study of SR in systems with a
density-dependent (diffusive-like) coupling. We initially discuss
the case of a scalar system \cite{extend3c}, and afterwards extent
the analysis to an array of FitzHugh--Nagumo units, with a
field-dependent activator diffusion \cite{TW}. For the second
system, when both diffusions are constant (that is: \(D_u>0\) and
\(D_v=0\)), has a known form of the NEP \cite{extend3b}. However,
in the general case we have not been able to find the form of the
NEP (but the idea of such a NEP is always \emph{underlying} our
analysis) and have to resort to an analysis based on numerical
simulations. The result shows that the system's response is
enhanced due to the particular form of the non-homogeneous
coupling. From such results, we can conclude that the phenomenon of
enhancement of the SNR, due to a selectivity in the coupling,
initially found for a scalar system \cite{extend3c} is robust, and
that the indicated non-homogeneous coupling could clearly
contribute to enhance the SR phenomenon in very general systems.

Finally, we analyzed an activator-like field including a nonlocal
contribution that arise through an effective adiabatic elimination
of an auxiliary (inhibitor-like) field. By exploiting the knowledge
of the nonequilibrium potential in such a case, we have analyzed
the dependence of the SNR on the nonlocal interaction kernel range,
founding that there is an optimal value of the kernel range,
yielding a maximum in the system's response, corresponding to a
very localized interaction.

The indicated results clearly show that the ``nonequilibrium
potential'' (even if not known in detail \cite{quasi}) offers a
very useful framework to analyze a wide spectrum of characteristics
associated to SR in spatially extended or coupled systems. For
instance, within this framework, the phenomenon of SSSR looks---as
other aspects of SR in extended systems \cite{extend3a}---as a
natural consequence of a breaking of the symmetry of the NEP
\cite{SSSR4}.
\section*{Appendix: Brief review of the nonequilibrium potential
scheme}
Loosely speaking, the notion of NEP is an extension to
nonequilibrium situations of that of equilibrium thermodynamic
potential. In order to introduce it, we consider a general system
of nonlinear stochastic equations (admitting the possibility of
\emph{multiplicative noises})
\begin{equation}\label{eq:14}
\dot{q}^\nu=K^\nu(q)+g^\nu_i(q)\,\xi_i(t),\qquad\nu=1,\ldots,n;
\end{equation}
where repeated indices are summed over. Equation (\ref{eq:14}) is
stated in the sense of It\^o. The \(\{\xi_i(t)\}\),
\(i=1,\ldots,m\leq n\) are mutually independent sources of Gaussian
white noise with typical strength \(\gamma\).
\subsection*{Graham's approach}
The Fokker--Planck equation corresponding to Eq.\ (\ref{eq:14})
takes the form
\begin{equation}\label{eq:15}
\frac{\partial P}{\partial t}=-\frac{\partial}{\partial q^\nu}
\,K^\nu(q)\,P+\frac{\gamma}{2}\frac{\partial^2}{\partial q^\nu
\,\partial q^\mu}\,Q^{\nu\mu}(q)\,P
\end{equation}
where \(P(q,t;\gamma)\) is the probability density of observing
\(q=(q_1,\ldots,q_n)\) at time \(t\) for noise intensity \(\gamma\),
and \(Q^{\nu\mu}(q)=g^\nu_i(q)\,g^\mu_i(q)\) is the matrix
of transport coefficients of the system, which is symmetric and
non-negative. In the long time limit (\(t\to\infty\)), the solution
of Eq.\ (\ref{eq:15}) tends to the stationary distribution
\(P_\mathrm{st}(q)\). According to \cite{GR}, the NEP
\(\Phi(q)\) associated to Eq.\ (\ref{eq:15}) is defined by
\begin{equation}\label{eq:16}
\Phi(q)=-\lim_{\gamma\to0}\gamma\,\ln P_\mathrm{st}(q,\gamma).
\end{equation}
In other words \[P_\mathrm{st}(q)\,d^nq=Z(q)\exp
\left[-\frac{\Phi(q)}{\gamma}+\mathcal{O}(\gamma)\right]\,d\Omega_q,\]
where \(\Phi(q)\) is the NEP of the system and \(Z(q)\) is defined
as the limit \[\ln Z(q)=\lim_{\gamma\to0}
\left[\ln P_\mathrm{st}(q,\gamma)+\frac{1}{\gamma}\,\Phi(q)\right].\]
Here \(d\Omega_q=d^nq/\sqrt{G(q)}\) is the invariant volume element
in the \(q\)-space and \(G(q)\) is the determinant of the
contravariant metric tensor (for the Euclidean metric it is
\(G=1\)). It was shown \cite{GR} that \(\Phi(q)\) is the solution
of a Hamilton--Jacobi-like equation (HJE)
\[K^\nu(q)\frac{\partial\Phi}{\partial q^\nu}+
\frac{1}{2}Q^{\nu\mu}(q)\frac{\partial\Phi}{\partial
q^\nu}\frac{\partial\Phi}{\partial q^\mu}=0,\] and \(Z(q)\) is
the solution of a linear first-order partial differential equation
depending on \(\Phi(q)\) (not shown here).

Equation (\ref{eq:16}) and the normalization condition ensure that
\(\Phi\) is bounded from below. Furthermore, from the separation of
the streaming velocity of the probability flow in the steady state
into conservative and dissipative parts, it follows that
\[\frac{d\Phi(q)}{dt}=K^\nu(q)
\frac{\partial\Phi(q)}{\partial q^\nu}=
-\frac{1}{2}\,Q^{\nu\mu}(q)\,\frac{\partial\Phi}
{\partial q^\nu}\frac{\partial\Phi}{\partial q^\mu}\leq 0,\]
i.e. \(\Phi\) is a LF for the dynamics of the system when
fluctuations are neglected. Under the deterministic dynamics,
\(\dot{q}^\nu=K^\nu(q)\), \(\Phi\) decreases monotonically and
takes a minimum value on attractors. In particular, \(\Phi\) must
be constant on all extended attractors (such as limit cycles or
strange attractors) \cite{GR}.
\subsection*{Ao's approach}
An alternative way to look into this problem is due to Ao
\cite{pao}. Let us refer again to the system in Eq.\ (\ref{eq:14}).
Following \cite{pao}, we introduce now the auxiliary matrix
\[\mathbf{H}^{-1}(\vec{q})=\mathbf{S}(\vec{q})+\mathbf{A}(\vec{q}),\]
with \(\mathbf{S}(\vec{q})\) a symmetric matrix while
\(\mathbf{A}(\vec{a})\) is an antisymmetric one.
\(\mathbf{H}^{-1}(\vec{q})\) is now used to rewrite the initial
system as
\[\mathbf{H}^{-1}(\vec{q})\,\dot{\vec{q}}=
\mathbf{H}^{-1}(\vec{q})\,\vec{K}(\vec{q},t)+\mathbf{H}^{-1}(\vec{q})\,
\vec{\xi}(\vec{q},t)=-\nabla\Phi(\vec{q},t)+\vec{\eta}(\vec{q},t),\]
where
\(-\nabla\Phi(\vec{q},t)=\mathbf{H}^{-1}(\vec{q})\,\vec{K}(\vec{q},t)\),
\(\vec{\eta}(\vec{q},t)=\mathbf{H}^{-1}(\vec{q})\,\vec{\xi}(\vec{q},t)\)
and
\(\mathbf{H}^{-1}(\vec{q})\,\dot{\vec{q}}=-\nabla\Phi(\vec{q},t)+
\vec{\eta}(\vec{q},t)\).
The new stochastic variables, \(\vec{\eta}(\vec{q},t)\), fulfill
\[\langle\vec{\eta}(\vec{q},t)\vec{\eta}^T(\vec{q},t')\rangle=2\,
\mathbf{S}(\vec{q})\,\delta(t-t')=2\,\mathbf{G}^{-1}(\vec{q})
\mathbf{Q}(\vec{q})[\mathbf{G}^{-1}(\vec{q})]^T\delta(t-t'),\]
that imposes a condition on the arbitrary definition of
\(\mathbf{S}(\vec{q})\) as we have
\[[\mathbf{S}(\vec{q})+\mathbf{A}(\vec{q})]\,\mathbf{Q}(\vec{q})\,
[\mathbf{S}(\vec{q})-\mathbf{A}(\vec{q})]=\mathbf{S}(\vec{q}).\]

As shown by Ao, the last equation also implies
\(\mathbf{H}(\vec{q})+\mathbf{H}^T(\vec{q})=
2\,\mathbf{Q}(\vec{q})\). We also have
\(\nabla\times\left[\mathbf{H}^{-1}(\vec{q})\,\vec{K}(\vec{q},t)
\right]=\nabla\times\left[-\nabla\Phi(\vec{q},t)\right]=0\). From
the previous equations, we have in principle all the needed
conditions to determine \(\mathbf{H}^{-1}(\vec{q})\), and to obtain
from it the potential \(\Phi(\vec{q},t)\). The interesting feature
of this approach is that it resorts neither to
\(P_\mathrm{st}(q)\) nor to the small-noise limit, thus being
applicable in principle to more general situations.
\section*{Acknowledgements}
The authors acknowledge the collaboration of B. von Haeften, S.
Bouzat, C. J. Tessone, G. G. Iz\'us, M. Kuperman, S. Mangioni, A.
S\'anchez, F. Castelpoggi, in different aspects and/or stages of
this work. HSW thanks the European Commission for the award of a
\emph{Marie Curie Chair} at the Universidad de Cantabria, Spain.


\begin{thebibliography}{99}
\bibitem{comment} It is worthwhile noting that when the parameter
\(h\) is large enough, under some circumstances the coupling term
might become negative, in what is known as ``inhibitory coupling''
\cite{Dayan}. This is a very interesting kind of coupling that has
attracted much attention, both in neural and chemical context, that
we will not discuss here.
\end{thebibliography}
\end{document}